\documentclass[apj]{emulateapj}
\usepackage{color}
\usepackage[linktocpage]{hyperref}
\usepackage{graphicx,amssymb,amsmath,amsthm,amsfonts,epsfig,bm}

\newcommand{\tn}[1]{{\rm #1}}
\newcommand{\nn}{\nonumber}

\begin{document}

\title{Geodesic models of quasi-periodic-oscillations as probes of quadratic gravity}

\author{Andrea Maselli\altaffilmark{1}, Paolo Pani\altaffilmark{2,3,4}, Roberto Cotesta\altaffilmark{5}, Leonardo Gualtieri\altaffilmark{2,3}, Valeria Ferrari \altaffilmark{2,3}, and Luigi Stella\altaffilmark{6}}

\altaffiltext{1}{Theoretical Astrophysics, Eberhard Karls University of Tuebingen, 
Tuebingen 72076, Germany}
\altaffiltext{2}{Dipartimento di Fisica, ``Sapienza'' Universit\`a di Roma, Piazzale Aldo Moro 5, 00185, Roma, Italy.}
\altaffiltext{3}{Sezione INFN Roma1, Piazzale Aldo Moro 5, 00185, Roma, Italy.}
\altaffiltext{4}{CENTRA, Departamento de F\'{\i}sica, Instituto Superior T\'ecnico,  Universidade de Lisboa, Avenida~Rovisco Pais 1, 1049 Lisboa, Portugal.}

\altaffiltext{5}{Max Planck Institute for Gravitational Physics (Albert Einstein Institute), Am Mühlenberg 1, Potsdam 14476, Germany.}

\altaffiltext{6}{INAF-Osservatorio Astronomico di Roma, via Frascati 33, 00078, Monteporzio Catone, Roma, Italy.}

\begin{abstract}
Future very-large-area X-ray instruments (for which the effective area is larger than $>3$~m$^2$) will be able to measure 
the frequencies of quasi-periodic oscillations~(QPOs) observed in the X-ray flux from accreting compact 
objects with sub-percent precision. If correctly modeled, QPOs can provide a novel way to test the strong-field 
regime of gravity.
By using the relativistic precession model and a modified version of the epicyclic resonance model, we develop 
a method to test general relativity against a generic class of theories with quadratic curvature corrections. With 
the instrumentation being studied for future missions such as eXTP, LOFT, or STROBE-X, a measurement of at 
least two QPO triplets from a stellar mass black hole  can set stringent constraints on the coupling parameters 
of quadratic gravity.
\end{abstract}


%%%%%%%%%%%%%%%%%%%%%%%%%%%%%%%%%%%%%%%%%%%%
\section{Introduction}\label{intro}
%%%%%%%%%%%%%%%%%%%%%%%%%%%%%%%%%%%%%%%%%%%%

The study of gravity near compact objects is among the last missing pieces of the grand program aimed 
at testing general relativity (GR) at all sub-galactic scales (for a recent review, see~\citet{0264-9381-32-24-243001}). 
Because of their simplicity, black holes~(BHs) are particularly well suited for testing gravity in the 
strong-field regime, which characterizes the dynamics near the horizon (for a recent review of BH-based 
tests of gravity, see~\citet{Yagi:2016jml}). X-rays emitted by matter accreting into stellar mass BHs
provide a very promising probe of the inner region of the accretion disk, which is believed to be bounded 
by the Innermost Stable Circular Orbit (ISCO) of the BH. In this region, the gravitational field cannot 
be described by Newtonian theory or by a weak-field expansion of GR: the strong-field 
regime of gravity is manifest there and there can be tested there.

Some X-ray spectroscopy features have been used as diagnostic tools of the inner disk region of BH accretion. 
These comprise the soft X-ray continuum emission, from which estimates of the ISCO location, and thus the 
BH rotation rate have been obtained~\citep{McClintock:2011} and the broad iron K$\alpha$ line and reflection 
spectrum, whose shape carries information on a variety of GR effects in the inner disk region~\citep{fabian_2012}. 
Among timing diagnostics, the multiple Quasi-periodic Oscillations (QPOs), which occur simultaneously in the X-ray 
flux from accreting stellar mass BHs and neutron stars~\citep{2006csxs.book...39V} are especially promising
\footnote{The potential of combined spectral-timing measurements on timescales comparable to the dynamical 
timescales of the inner disk regions is currently being investigated through some GR-based studies, while modeling of 
presently available X-ray measurements has already provided interesting results (see e.g. \citep{Uttley:2014zca}).}.
Most QPO models, including the two models we 
adopt here, {\it i.e.} the relativistic precession model (RPM; \citep{Stella:1999sj}) and the epicyclic 
resonance model (ERM; \citep{2005A&A...436....1T}), involve frequencies associated to the orbital motion of matter in 
the inner disk, which is directly determined by the characteristics of the strong gravitational field in this region. 
Both the RPM and (a recently introduced extension of) the ERM \citep{Stuchlik:2016kqo} aim at interpreting 
three QPO signals that have been observed simultaneously in a number 
of accreting neutron star systems and, so far, in only one accreting BH system, GRO J1655-40.
These three signals comprise: (1) a low frequency (LF) QPO at $\nu_{\rm LF}$, which is the so-called 
type~C QPO in BH systems and Horizontal Branch QPO in neutron star systems \citep{Casella:2005vy},
with frequencies of up to tens of Hz and (2) twin high frequency (HF) QPOs, at $\nu_\tn{lower}$ and $\nu_\tn{upper}$, 
with frequencies of several hundred Hz in BHs and around $\sim 1$~kHz NSs. 
Since these QPOs are detected as incoherent signals in the power spectra of high-time resolution X-ray light curves, 
their signal-to-noise ratio (S/N) scales linearly with the source count rate and thus with the effective area of X-ray 
instrumentation. Most currently available QPO measurements have been obtained with the Proportional Counter 
Array instrument on board the Rossi X-ray Timing Explorer (RXTE/PCA). 
By exploiting monolithic Silicon Drift Detector technology \citep{2010SPIE.7732E..1VF}
the next generation X-ray astronomy satellites, which are currently being studied, including 
LOFT \citep{2016SPIE.9905E..1RF}, eXTP \citep{2016SPIE.9905E..1QZ} and STROBE-X \citep{2017AAS...22930904W}
will achieve an order of magnitude increase in effective area with respect to RXTE/PCA and thus obtain 
high-precision measurements of simultaneous QPO signals from a variety of BH systems will then become 
possible. In this paper, we investigate the way in which QPO as measured with the eXTP 
Large Area Detector (eXTP/LAD, factor of $\sim 6$ larger area than RXTE/PCA)  
may afford testing the strong-field/high-curvature regime of GR against some 
alternative theories. 

Modified gravity theories can be introduced either by using a bottom-up approach, in which one considers 
phenomenological parametrizations of  BH spacetimes (or of other observable quantities) depending
on a set of parameters, or by using a top-down approach, in which  specific modifications of GR, possibly 
inspired by fundamental physics considerations, are adopted~\citep{Psaltis:2009xf}. No practical and 
sufficiently general parametrization of  deviations from GR in the strong-field regime 
has yet been proposed: therefore, we shall follow a top-down approach. Since we are interested in 
testing the strong-field/large-curvature regime of gravity, we shall consider 
the so-called ``quadratic gravity theories'', which are the simplest and most natural modifications 
of GR in this regime.

In quadratic gravity theories, the Einstein-Hilbert action is modified by including quadratic terms in the curvature
tensor, coupled with a scalar field. These couplings can be interpreted as the first terms in an expansion taking into account
all possible curvature invariants; such expansion (which is suggested by low-energy effective string
theories~\citep{Gross:1986mw}) could make the theory renormalizable~\citep{Stelle:1976gc}. 

The action (in vacuum), which includes all quadratic curvature invariants, generically coupled to a single scalar field, 
can be written as (see for instance \citep{Yunes:2011we,0264-9381-32-24-243001} and references therein)
\begin{eqnarray}
 S&&=\frac{1}{16\pi}\int\sqrt{-g} d^4x \left[R-\frac{1}{2}\nabla_a\phi\nabla^a\phi+f_1(\phi)R^2\right.\nn\\
&&\left.+f_2(\phi) R_{ab}R^{ab}+f_3(\phi) R_{abcd}R^{abcd}+f_4(\phi)^*RR\right] \,, \nn \\\label{action}
\end{eqnarray}
where  $f_i(\phi)$ ($i=1, 2, 3, 4$) are generic coupling functions, ${}^{*}\!RR\equiv\frac{1}{2}R_{\mu\nu\rho\sigma}\epsilon^{\nu\mu\lambda\kappa}R^{\rho\sigma}{}_{\lambda\kappa}$, with $\epsilon^{\mu\nu\rho\sigma}$ the Levi-Civita tensor. Two relevant cases of this class of theories are (1) Einstein-dilaton-Gauss-Bonnet (EDGB) gravity ($f_1=\frac{\alpha}{4} e^{\phi}$, $f_2=-4f_1$, $f_3=f_1$, $f_4=0$), in which the quadratic corrections reduce to the Gauss-Bonnet invariant, $R_{\rm GB}^2\equiv R^2-4R_{\mu\nu}^2+R_{\mu\nu\rho\sigma}^2$~\citep{Kanti:1995vq,Moura:2006pz}; and (2) Dynamical Chern-Simons (DCS) gravity (\citet{Jackiw:2003pm,Alexander:2009tp, Delsate:2014hba}; $f_1=f_2=f_3=0$ and $f_4=\frac{\beta}{4}\phi$). 

In general, the equations of motion of the action (\ref{action}) have third-(or higher-)order derivatives, and the
theory is subject to Ostrogradsky's instability~\citep{Woodard:2006nt}. To avoid this feature, the theory should be treated as an
effective field-theory, valid only up to second order in the curvature, in the limit of small couplings $f_i$. In this
way, ghosts and other pathologies disappear (for a discussion, see~\citet{0264-9381-32-24-243001}). This limit also requires us 
to expand the functions $f_i(\phi)$ up to linear order in $\phi$, i.e. $f_i(\phi)\approx \eta_i+\frac{\alpha_i}{4}\phi$,
with $\eta_i,\alpha_i$ such that the corrections are small compared to the leading Einstein-Hilbert term~\citep{Yunes:2011we,Pani:2011gy,0264-9381-32-24-243001}. The only exception is EDGB gravity, whose
equations of motion are second order, avoiding Ostrogradsky's instability. 
Therefore, EDGB gravity can be treated as an
``exact'' theory, and its coupling constant $\alpha$ can, in principle, be a finite quantity.

BH solutions to theories based on action~\eqref{action} have been found in various particular cases 
by~\citet{Mignemi:1992nt,Kanti:1995vq,Pani:2009wy,Yunes:2009hc,Yunes:2011we,Kleihaus:2011tg,Yagi:2012ya,Kleihaus:2014lba}.
Stationary, axisymmetric, BH solutions can be found in closed analytical form to any order in a small-spin and
small-coupling expansion~\citep{Pani:2011gy,Ayzenberg:2014aka,Maselli:2015tta}. 
Remarkably, to leading order in the coupling, the metric
depends only on two constants, 
$\alpha_3 \equiv \alpha_{\rm GB}=\alpha$ and $\alpha_4\equiv \alpha_{\rm DCS}=\beta$, 
whereas it is independent of $\eta_i$ ($i=1,2,3,4$) and of $\alpha_{1,2}$.
Thus, all stationary solutions to the effective-field theory introduced above reduce, to second 
order in the coupling parameter, to two families, namely BH
solutions to EDGB gravity and to DCS gravity (or, at most, solutions to theories with different choices of 
$f_3(\phi)$, $f_4(\phi)$; we remark that these theories are equivalent to EDGB and DCS gravity in the small-coupling limit).
The former case is the only one that can be defined beyond the small-coupling approximation; BH solutions in this 
case were obtained numerically for generic spin and coupling by~\citet{Kleihaus:2011tg,Kleihaus:2014lba} and in closed 
analytical form to fifth order in the spin and to seventh order in the coupling parameter by~\citet{Maselli:2015tta}. 
In the latter case, spinning DCS BHs were first obtained by~\citet{Yunes:2009hc,Konno:2009kg} to leading order in the 
spin and by~\citet{Yagi:2012ya} to quadratic order in the spin.

In this article, we discuss the possibility of using QPOs as tools to test GR against quadratic gravity theories,
extending previous results from~\citet{Vincent:2013uea} and \citet{Maselli:2014fca,Maselli:2015tta}. It is worth 
remarking that the QPO diagnostics has also been exploited in \citet{Bambi:2013fea,Bambi:2012pa}, and in
\citet{Franchini:2016yvq} as a probe of hairy BH 
solutions that emerge in standard GR in the presence of ultralight fields.

This paper is organized as follows. 
In Section~\ref{sec:epyc} we work out the characteristic orbital frequencies - i.e., 
the azimuthal and epicyclic frequencies - for rotating BHs in EDGB gravity and DCS gravity. 
Since DCS gravity has to be considered as an effective theory, in our analysis, we assume $\beta$ 
to be a small quantity, while we allow for finite values of the EDGB coupling parameter, though a 
theoretical bound  does exist for $\alpha$, i.e.
\begin{equation}\label{EDGBconst}
0<\frac{\alpha}{M^2}\lesssim 0.691\ ,
\end{equation}
where $M$ is the BH mass.

In this paper, we shall employ the BH solution in EDGB gravity derived
in~\citet{Maselli:2015tta} to fifth order in the spin and to seventh order in $\alpha$, and the BH solution in DCS gravity 
to fifth order in the spin (thus extending the results of~\citet{Yagi:2012ya}) and to second order in the coupling parameter $\beta$.  For the latter, the explicit expressions of the 
metric tensor and the scalar field are quite long and are available in a {\scshape Mathematica} notebook provided in the 
Supplemental Material. 
In Section~\ref{sec:models} we discuss the QPO models we adopt in our analysis, in which the
observed frequencies are related to the azimuthal and epicyclic frequencies of BH geodesics. In
Section~\ref{sec:analysis} we discuss the method we propose in order to test GR 
against EDGB gravity and DCS gravity by using future high-precision QPO measurements. 
We show that, using this approach,  large-area X-ray instrumentation such as that being  
studied for LOFT-P, eXTP, and STROBE-X can significantly constrain the parameter space of these theories.

We use geometric units in which $G=c=1$ and consider a spin parameter $a^\star=J/M^2<0.5$, such that 
truncation errors are expected to be of ${\cal O}(a^{\star 6})\approx 1\%$ at the most.

%%%%%%%%%%%%%%%%%%%%%%%%%%%%%%%%%%%%%%%%%%%%
\section{The epicyclic frequencies of a rotating BH}\label{sec:epyc}
%%%%%%%%%%%%%%%%%%%%%%%%%%%%%%%%%%%%%%%%%%%%

\begin{figure}[ht]
\centering
\includegraphics[width=8cm]{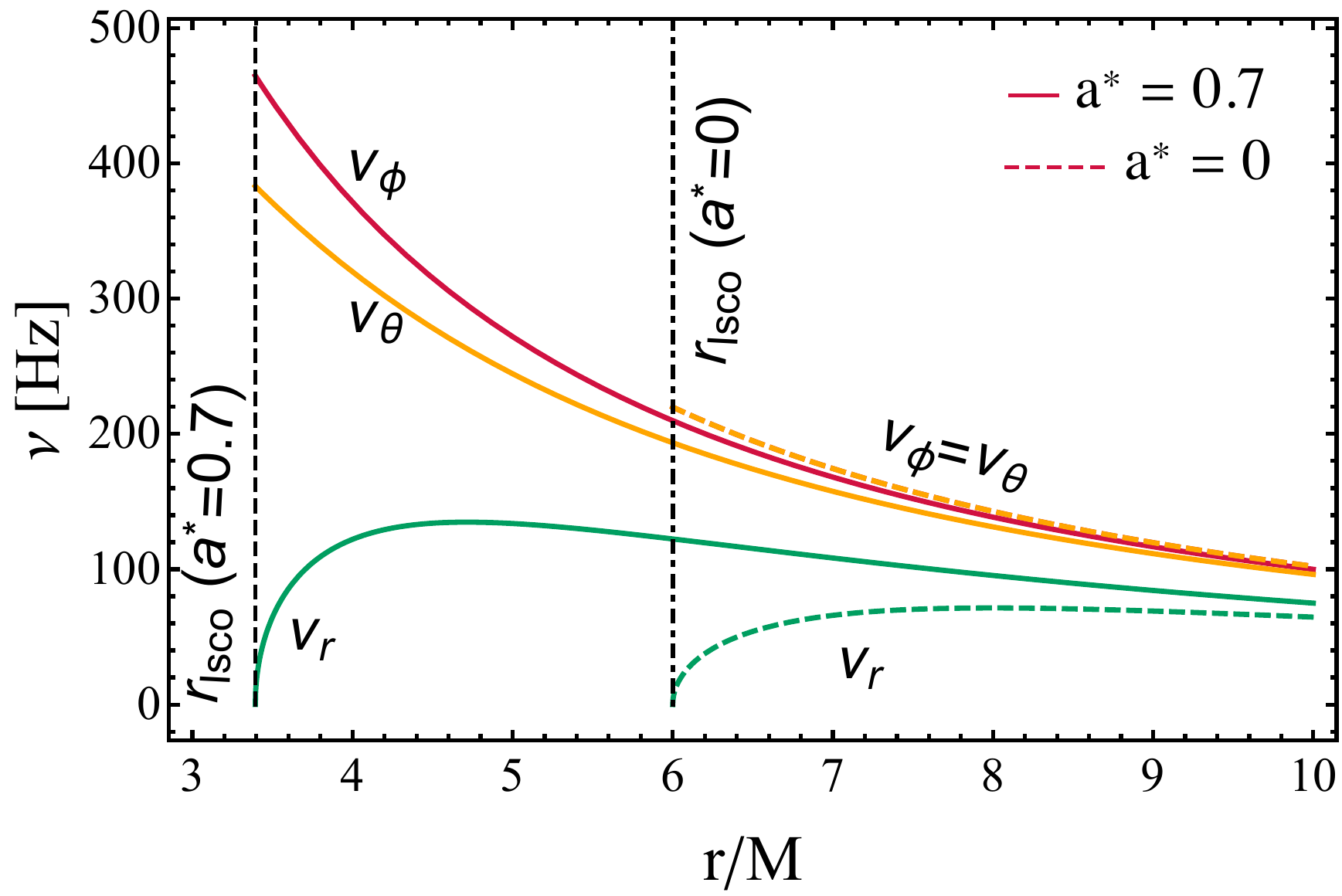}
\caption{Epicyclic frequencies for a Kerr BH with $M=10M_\odot$ as function of the 
dimensionless radial variable $r/M$ for two values of the spin parameter $a^\star=0$ 
(dashed curves) and $a^\star=0.7$ (solid curves). The dashed and dotted-dashed vertical lines 
represent the radius of the ISCO for the two considered cases.}
\label{epyfreq}
\end{figure}

In thin accretion disks  around a rotating BH the stream of matter follows nearly 
equatorial $(\theta=\pi/2)$ and nearly circular geodesics at a specific radius $r_0$. 
For an axially symmetric spacetime described by the line element 
\begin{equation}
ds^2=g_{tt}dt^2+g_{rr}dr^2+g_{\theta\theta}d\theta^2+g_{t\phi}dt d\phi+g_{\phi\phi}d\phi^2\ ,
\end{equation}
small perturbations of the particle trajectories along the radial and vertical directions  
$r=r_0+\delta r$ and $\theta=\pi/2+\delta\theta$, lead to oscillations around the 
equilibrium configuration characterized by the two epicyclic frequencies 
\begin{eqnarray}
\nu_r^2&=&\frac{1}{(2\pi)^2}\frac{(g_{tt}+\Omega g_{t\phi})^2}{2g_{rr}}\frac{\partial^2 {\cal U}}{\partial r^2}\left(r_0,\frac{\pi}{2}\right)\ ,\label{nur}\\
\nu_\theta^2&=&\frac{1}{(2\pi)^2}\frac{(g_{tt}+\Omega g_{t\phi})^2}{2g_{\theta\theta}}\frac{\partial^2 {\cal U}}{\partial \theta^2}\left(r_0,\frac{\pi}{2}\right)\label{nutheta}\ ,
\end{eqnarray}
where $\Omega=2\pi \nu_\phi$ is the particle angular velocity and $\nu_\phi$ its 
azimuthal frequency (see \citep{Maselli:2014fca} for technical details). The effective potential 
${\cal U}(r,\theta)=g^{tt}-2 l g^{t\phi}+l^2g^{\phi\phi}$ depends on the metric functions and 
the ratio between the particle's angular momentum and energy $l=L/E$ per unit mass. For a 
specific radius $r$, the three frequencies of a Kerr BH $(\nu_\phi,\nu_\theta,\nu_r)$ are functions 
of the mass and spin parameter $a^\star=J/M^2\in[-1,1]$ only, where $J$ is the BH angular 
momentum:
%%%
\begin{eqnarray}
\nu_\phi^{\tn{GR}}&=&\frac{1}{2\pi}\frac{M^{1/2}}{r^{3/2}+ {a^\star}
M^{3/2}}\,, \label{nu1GR}\\
\nu_r^{\tn{GR}}&=&\nu_\phi^{\tn{GR}}\left(1-\frac{6M}{r}+
8{a^\star}\frac{ M^{3/2}}{r^{3/2}}- 3{a^\star}^2\frac{
M^2}{r^2}\right)^{1/2}\,,\label{nu2GR}\\
\nu_\theta^{\tn{GR}}&=&\nu_\phi^{\tn{GR}}\left(1-4{a^\star}\frac{
M^{3/2}}{r^{3/2}}+ 3{a^\star}^2\frac{ M^2}{r^2}\right)^{1/2}\,.
\label{nu3GR} \end{eqnarray}
%%%
In Newtonian gravity, the three frequencies coincide, while in GR $\nu_\phi\geq \nu_\theta>\nu_r$. 
In particular, for $a^\star=0$ the azimuthal and vertical components are equal, whereas the 
radial one vanishes at the ISCO. As an example, in Fig.~\ref{epyfreq} we show these quantities for a BH with 
$M=10M_\odot$ and spin parameter $a^\star=(0,0.7)$. 

%%%%%%%%%%%%%%%%%%%%%%%%%%%%%%%%%%%%%%%%%%%%
%%%%%%%%%%%%%%%%%%%%%%%%%%%%%%%%%%%%%%%%%%%%

Epicyclic frequencies in the EDGB gravity have been computed in \citep{Maselli:2014fca} 
for slowly rotating BHs  at the linear order in the angular momentum, as a function 
of the coupling constant $\alpha/M^2$; they show differences with respect to the GR case, 
which increase for higher BH spin, and are potentially observable by future 
large-area X-ray satellites. Motivated by these results,~\citep{Maselli:2015tta} improved the templates for 
$(\nu_\phi,\nu_\theta,\nu_r)$ by extending them up to the fifth order in $a^\star$. 
Based on this higher-order expansion, more rapidly spinning BHs can be considered, thus
exploring regions in parameter space that are amenable 
to show significant departures from GR. 

\begin{figure}[ht]
\centering
\includegraphics[width=4.25cm]{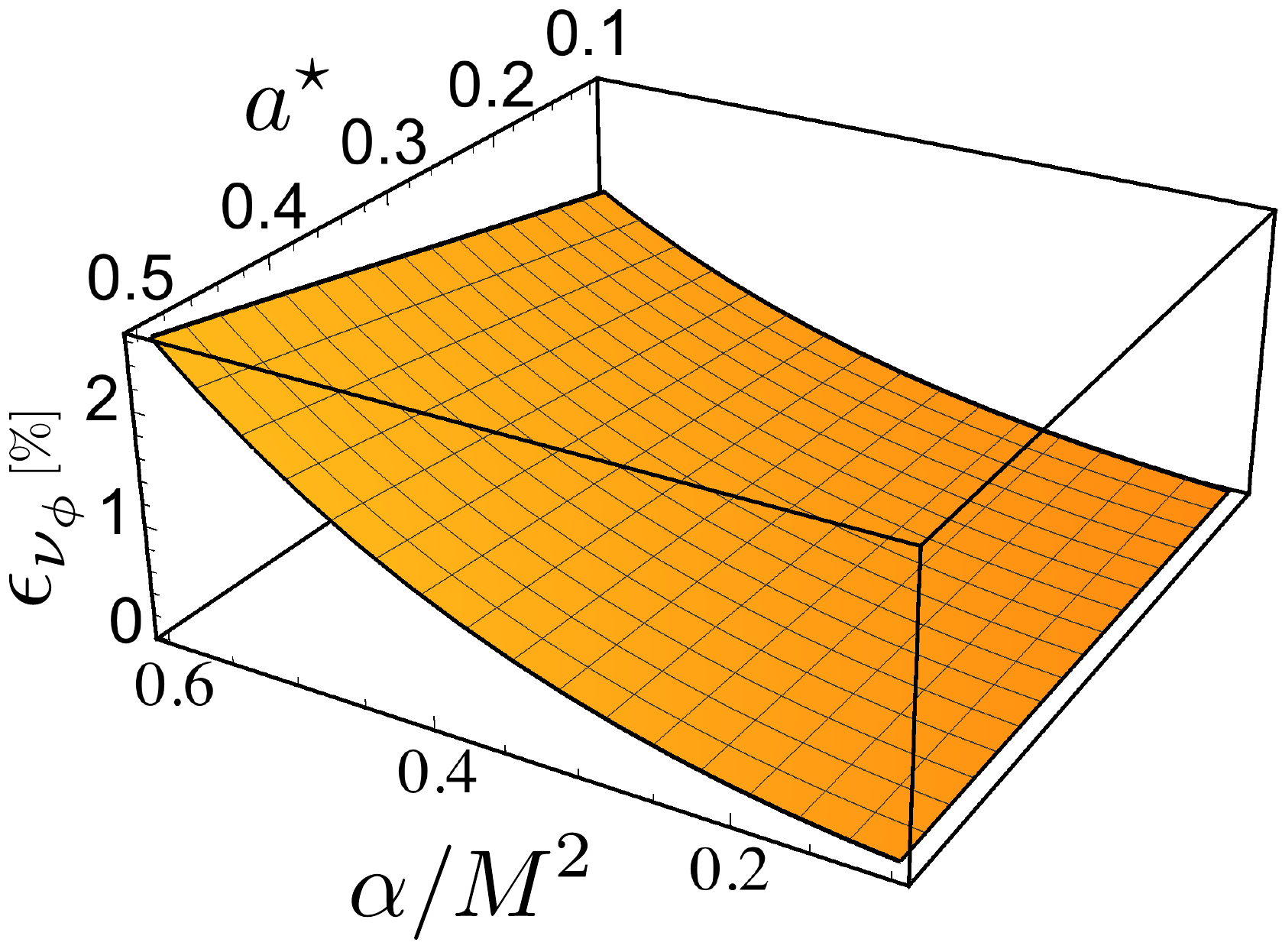}
\includegraphics[width=4.25cm]{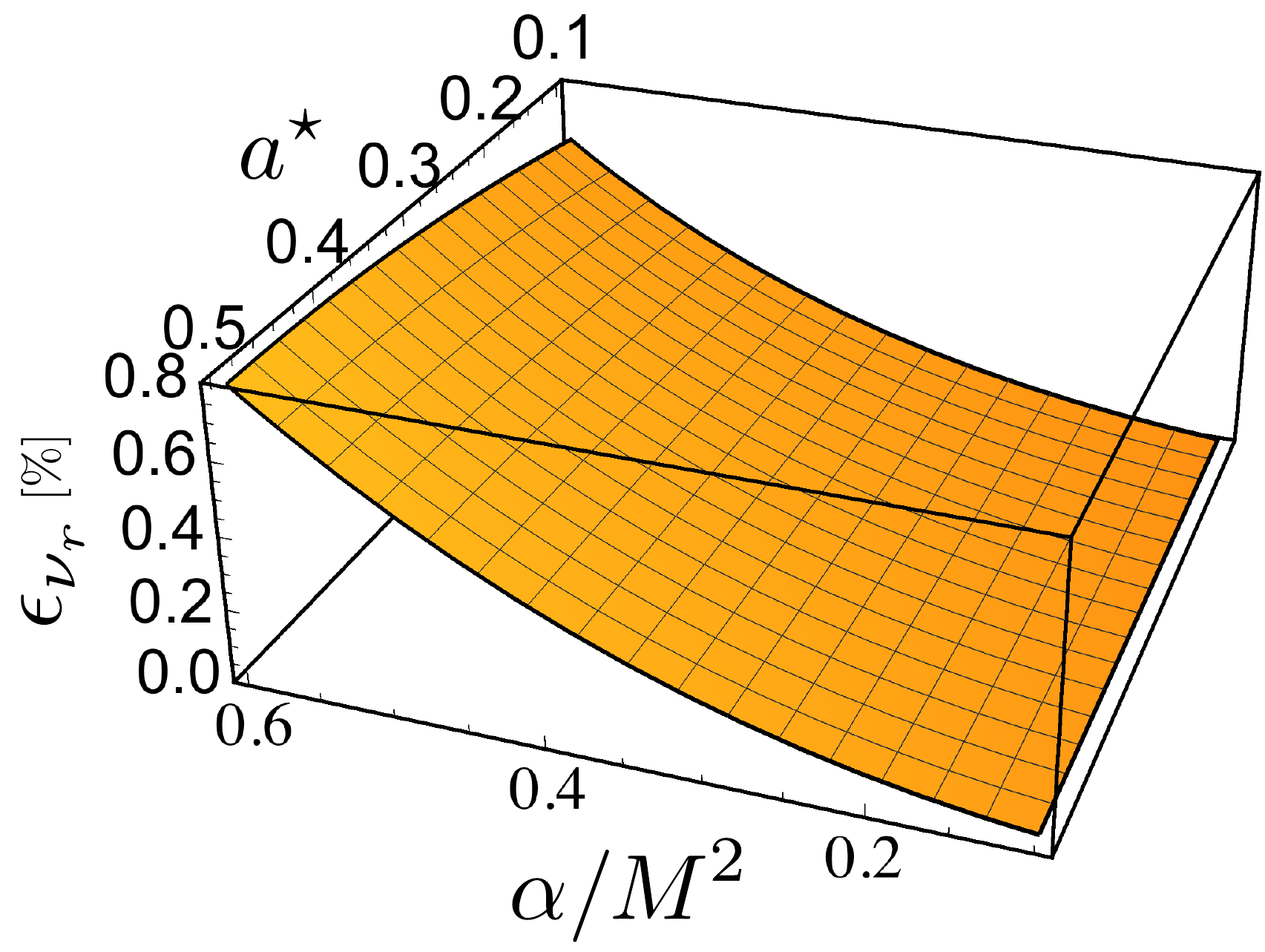}
\includegraphics[width=5cm]{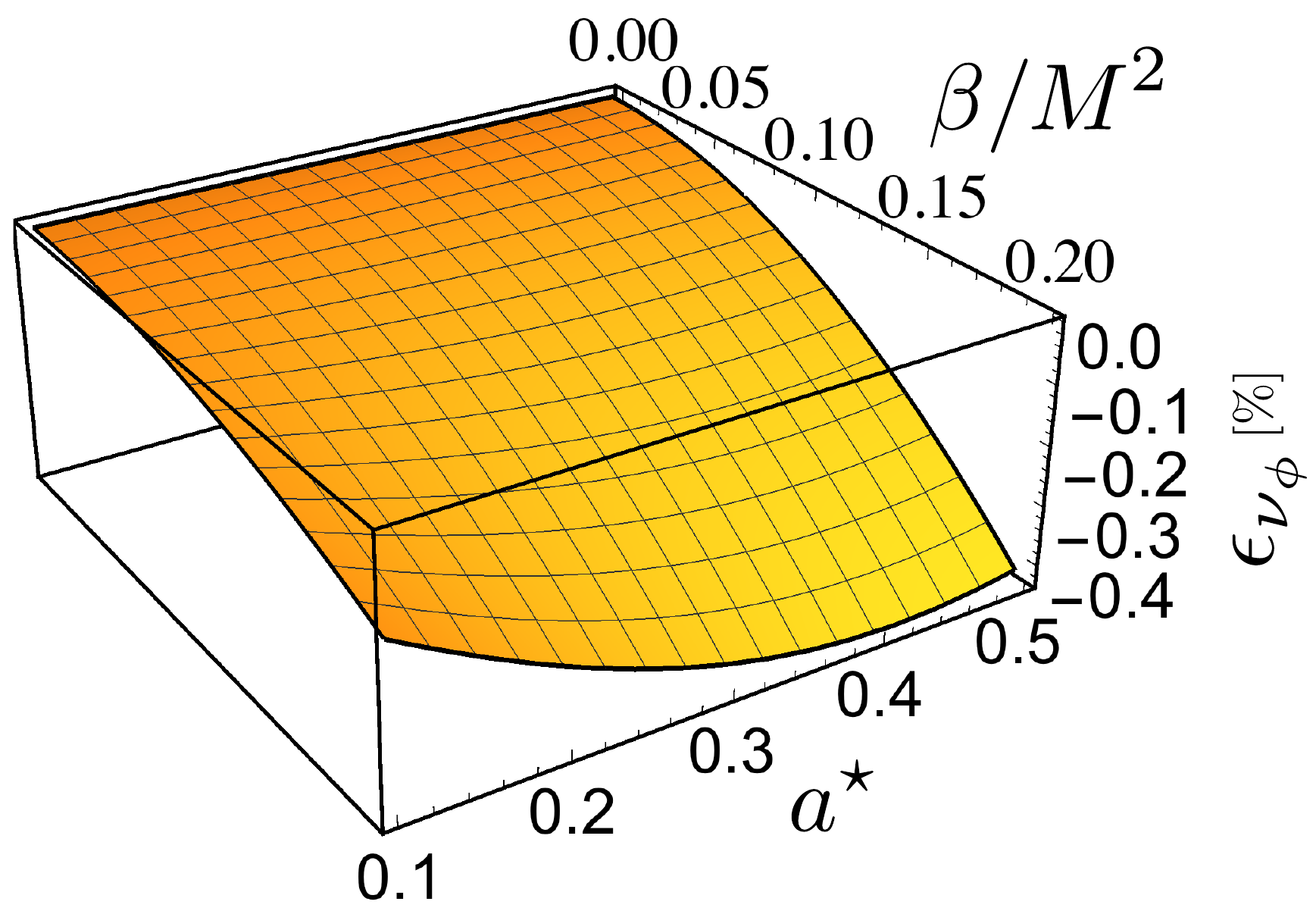}
\caption{Relative percentage difference between the values of  $(\nu_\phi,\nu_r)$ 
computed in GR and EDGB (top panels), and of $\nu_r$ derived in GR and DCS 
(bottom panel), assuming the emission radius at $r=1.1r_\tn{ISCO}$.}
\label{epyfreq2}
\end{figure}

To be consistent with the formalism developed for the EDGB and the DCS theories, in our 
analysis, we will expand Eqns.~(\ref{nu1GR})-(\ref{nu3GR}) as power series of $a^\star$ 
neglecting terms that are $\mathcal{O}(a^{\star 6})$ and higher. As an example, in Fig.~\ref{epyfreq2} 
we plot the relative percentage difference between the epicyclic frequencies $\nu_\phi$ and $\nu_r$ 
computed at $r=1.1r_\tn{ISCO}$ in GR and in EDGB or DCS theory:
\begin{equation}
\epsilon_{\nu_{\phi}}=\frac{\nu^\tn{EDGB,DCS}_{\phi}-\nu^\tn{GR}_{\phi}}{\nu^\tn{GR}_{\phi}}\ , 
\quad \epsilon_{\nu_r}=\frac{\nu^\tn{EDGB,DCS}_{\phi}-\nu^\tn{GR}_{r}}{\nu^\tn{GR}_{r}}\ ,
\end{equation}
as functions of the coupling parameters and of the BH spin. As expected, for larger values of 
$(\alpha/M^2,\beta/M^2)$ and faster rotation rates the relative difference increases; it can be as high 
as $3\%$ for the equatorial frequency in EDGB gravity. For the vertical component $\nu_\theta$ the 
relative difference is of the same order as that of $\nu_\phi$. These values decrease when the frequencies are 
computed in DCS. The bottom panel of Fig.~\ref{epyfreq2} shows that in this case $\epsilon_{\nu_{\phi}}\ll 1$ 
for all coupling parameter values allowed by the theory.

%%%%%%%%%%%%%%%%%%%%%%%%%%%%%%%%%%%%%%%%%%%%
%%%%%%%%%%%%%%%%%%%%%%%%%%%%%%%%%%%%%%%%%%%%

%%%%%%%%%%%%%%%%%%%%%%%%%%%%%%%%%%%%%%%%%%%% 
\section{Geodesic models of QPOs}\label{sec:models}
%%%%%%%%%%%%%%%%%%%%%%%%%%%%%%%%%%%%%%%%%%%% 

The epicyclic frequencies are the basic ingredients of the geodesic models that we shall 
describe in this section: the RPM \citep{Stella:1999sj}
and the modified ERM \citep{2005A&A...436....1T}. Both 
approaches  interpret the simultaneous occurrence of the LF and twin 
high frequency HF QPOs in terms of geodesic frequencies.

According to the RPM, the upper and lower HF QPOs coincide 
with the azimuthal frequency $\nu_\phi$, and the periastron
precession frequency, $\nu_\tn{per} = \nu_\phi-\nu_r$. 
The LF QPO mode instead, is identified with the nodal precession frequency, 
$\nu_\tn{nod}=\nu_\phi - \nu_\theta$. These three QPO signals 
$(\nu_\phi,\nu_\tn{per},\nu_\tn{nod})$ are assumed to be generated at the 
same orbital radius. Although the first application of the RPM to BH 
systems traced-back to the original paper\citep{Stella:1999sj}, 
the first complete exploitation of the
model has been made possible by the discovery of three simultaneous QPOs 
in GRO J1655-40, as measured by the 
RXTE/PCA with 0.5-1.5\% accuracy
\begin{equation}
\label{freq}
\nu_\phi= 441^{+2}_{-2}~\tn{Hz},\ 
\nu_\tn{per}= 298^{+4}_{-4}~\tn{Hz},\ 
\nu_\tn{nod}= 17.3^{+0.1}_{-0.1}~\tn{Hz}\ .
\end{equation}

By assuming the Kerr metric and fitting the three values in terms of the RPM frequencies, a 
precise estimate of the emission radius $r=(5.68 \,\pm 0.04)M$, and of the BH mass 
$M=(5.31\pm 0.07) M_\odot$ and angular momentum $a^\star=J/M^2 = 0.290 \pm 0.003$ 
was obtained \citep{Motta:2013wga}. Independent optical observations of the same source 
lead to mass values in good agreement with those obtained through the RPM 
$[M_\tn{opt}=(5.4\pm0.3)M\odot]$.  However, there exist systematic uncertainties in different methods 
for estimating BH spin, as spectral continuum measurements and analysis of the Fe spectral line profiles 
predict BH spin in the range $0.65<a^\star_\tn{sc}<0.75$ and $0.94<a^\star_\tn{Fe}<0.98$, respectively 
\citep{Motta:2013wwa}, which show a discrepancy with the QPO-based value and among each other.

The ERM builds on the possibility that in BHs with twin  HF QPOs, their centroid 
frequencies are such that $\nu_\tn{upper}/\nu_\tn{lower}=3/2$ \citep{2005A&A...436....1T}. 
This suggests an underlying mechanism based on nonlinear resonances. 
At the first order in the vertical and radial displacements, deviations $\delta r$ and 
$\delta \theta$ from geodesic circular motion can be described by the following equations:
\begin{equation}\label{reseq}
\ddot{\delta r}+\omega_{r}^2\delta_r=\delta a_r\quad \ ,\quad \ddot{\delta \theta}+\omega_{\theta}^2\delta_r=\delta a_\theta\ , 
\end{equation}
where dots represent time derivatives, $\omega_{\theta,r}=2\pi \nu_{r,\theta}$, and 
$\delta a_r, \delta a_\theta$ are forcing terms. Nonlinear resonances show interesting common features with 
QPOs, e.g.\ they occur only over a finite $\delta \nu$, allow for frequency combinations, and for sub-harmonic 
modes. 
If $\delta r \gg\delta \theta$, the mixing term $\delta\theta \delta r$ cannot be neglected and it 
must be included in the linearized equation for the vertical component (\ref{reseq}). 
For $\delta a_r=\delta a_\theta=0$, the radial component yields the solution $\delta_r=A \cos(\omega_r t)$, 
while the vertical oscillation obeys the Mathieu equation 
\begin{equation}
\ddot{\delta\theta}+\omega^2_\theta[1+ A h\cos(\omega_r t)]\delta \theta=0\ ,
\end{equation}
where $A$ and $h$ are known constants. The above equation describes parametric resonances 
such that 
\begin{equation}\label{resonancecon}
\frac{\nu_r}{\nu_\theta}=\frac{2}{n}\quad\ ,\quad n=1,2,3\ldots \ .
\end{equation}
In GR, $\nu_\theta>\nu_r$ and therefore we may associate the vertical component to the 
larger of the HF QPOs. It is interesting to note that a nonzero forcing term along the $\theta-$direction, 
i.e. $\delta a_\theta\neq0$, allows for a combination of the frequencies
\begin{equation}
\nu_-=\nu_\theta -\nu_r \quad \ ,\quad \nu_+=\nu_\theta +\nu_r\ ,
\end{equation}
which still satisfies the observational evidence for a small integer ratio 
$\nu_\tn{upper}/\nu_\tn{down}=3/2$, as long as
\begin{equation}
\nu_\tn{upper}=\nu_\theta\quad\ , \quad \nu_\tn{lower}=\nu_-\ ,
\end{equation}
or
\begin{equation}
\nu_\tn{upper}=\nu_+\quad\ , \quad \nu_\tn{lower}=\nu_\theta\ 
\end{equation}
are considered.
The ERM, originally developed to interpret only the twin HF QPOs, has been 
extended to interpret the LF QPO mode in terms of $\nu_\tn{nod}$ (as in the RPM),
and thus interpret the simultaneous occurrence of the three 
frequencies associated with GRO J1655-40 \citep{2016A&A...586A.130S}. Among all  
possibilities discussed by these authors, we consider here the case in which the two HF QPOs 
are identified with $\nu_\tn{upper}=\nu_\theta$ and $\nu_\tn{lower}=\nu_-$. 
With this choice, the BH parameters have been constrained to 
$M=(5.1\pm 0.1) M_\odot$, $a^\star= 0.274 \pm 0.003$ and 
$r=(5.67 \,\pm 0.05)M$, which are reasonably close to those derived with the RPM.
Similar results hold also for different combinations, since the fundamental scale of the 
effect is set by the ISCO frequency.

%%%%%%%%%%%%%%%%%%%%%%%%%%%%%%%%%%%%%%%%%%%%
\begin{figure}[ht]
\centering
\includegraphics[width=4.25cm]{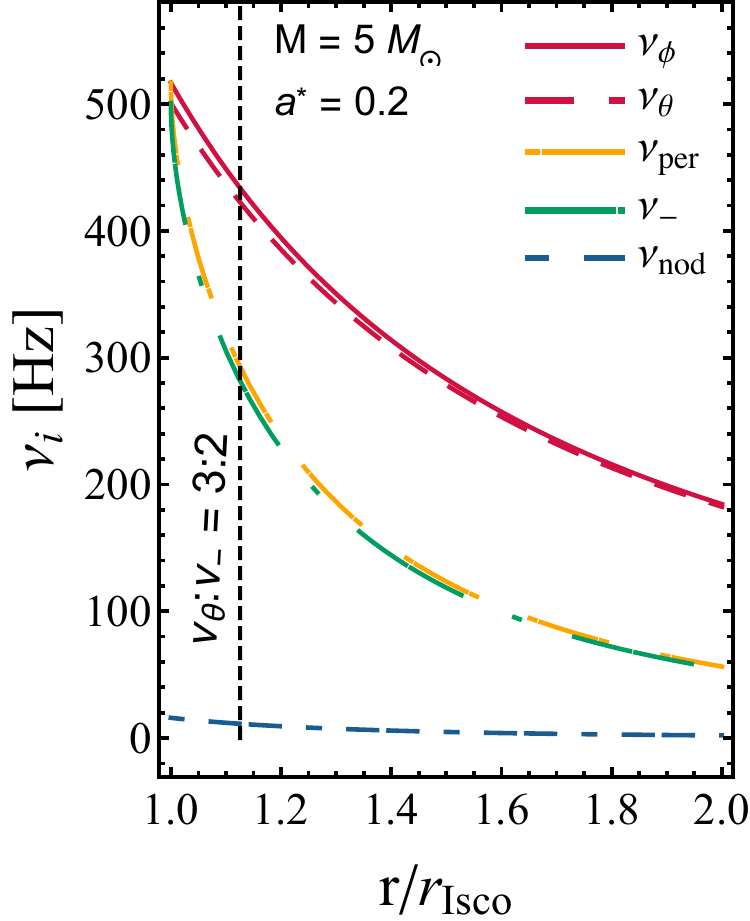}
\includegraphics[width=4.25cm]{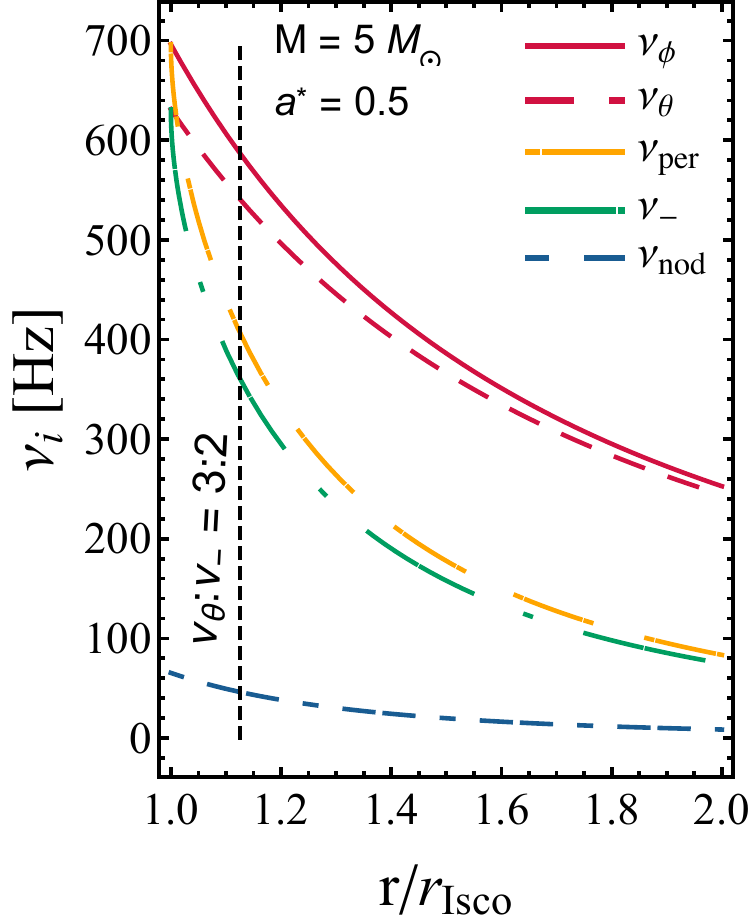}
\caption{We show the values of $(\nu_\phi,\nu_\theta,\nu_\tn{per},\nu_-,\nu_\tn{nod})$ for a Kerr BH with mass 
$M=5M_\odot$, $a^\star=0.2$ (left), and $a^\star=0.5$ (right), as a function of the radius normalized to the ISCO radius. 
The vertical dashed line corresponds to the orbital distance for which $\nu_\theta$ and $\nu_-$ are in resonance, with 
their ratio being $\nu_\theta/\nu_-=1.5$.}
\label{epyfreq3}
\end{figure}
%%%%%%%%%%%%%%%%%%%%%%%%%%%%%%%%%%%%%%%%%%%%

As an example, in Fig.~\ref{epyfreq3}, we show the two sets of frequencies $(\nu_\phi,\nu_\tn{per},\nu_\tn{nod})$ 
and $(\nu_\theta,\nu_\tn{nod},\nu_\tn{-})$ employed by the RPM and the ERM approaches. All values 
are computed in GR for a $5 M_\odot$ BH with different spin, as a function of radius in $r_\tn{Isco}$ units. 

%%%%%%%%%%%%%%%%%%%%%%%%%%%%%%%%%%%%%%%%%%%%
\section{QPOs and BHs in quadratic gravity}\label{sec:analysis}
%%%%%%%%%%%%%%%%%%%%%%%%%%%%%%%%%%%%%%%%%%%%

To test the ability of eXTP/LAD to constrain quadratic theories of gravity, we follow the same data analysis 
procedure described in \citep{Maselli:2014fca}, based on simulations of two 
QPO triplets with different values of their frequencies. In the following, we briefly summarize the basic steps of 
this method.

We consider a prototype BH of mass $\bar{M}=5.3M_\odot$ and selected values of its  
$\bar{a}^\star$, and adopt different values of the coupling parameter for each modified gravity theory. 
By using the analytic relations derived in \citep{Maselli:2015tta}, we calculate two 
sets of epicyclic frequencies for two different emission radii $r_{1,2}$, in both EDGB or DCS gravity.
Based on these sets, we calculate the three QPO signals expected in the case of the RPM and ERM; 
we then use the corresponding two QPO triplets as center values to generate $N=10^5$ samples 
from Gaussian distributions with standard deviations 
$(\sigma_{\varphi},\sigma_\tn{nod},\sigma_\tn{per})$,
obtained by rescaling the error bars in Eq. (9) by the ratio of the 
RXTE/PCA and eXTP/LAD effective areas.
Finally, we use the geodesic frequencies for a Kerr BH (Equations~(\ref{nu1GR})-(\ref{nu3GR}))
to compute $2N$ triplets of $(M,a^\star,r)_{j=1,2}$, from which 
the mean values of the BH mass, spin, and emission radii, and the 
covariance matrices $\Sigma_1,\Sigma_2$ associated with the two sets are derived. 

%%%%%%%%%%%%%%%%%%%%%%%%%%%%%%%%%%%%%%%%%%%%
\begin{figure*}[ht]
\centering
\includegraphics[width=5.3cm]{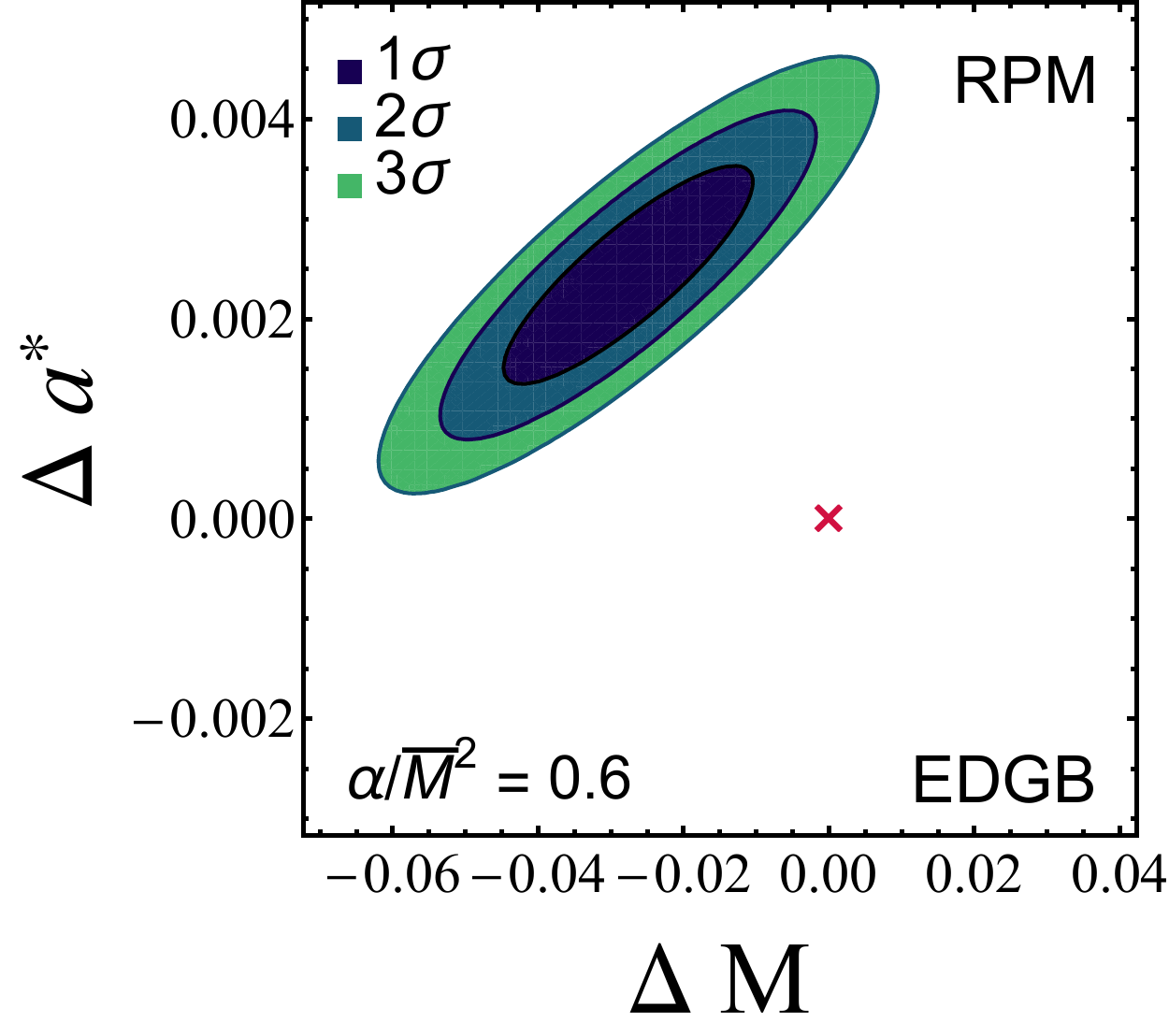}
\includegraphics[width=5.3cm]{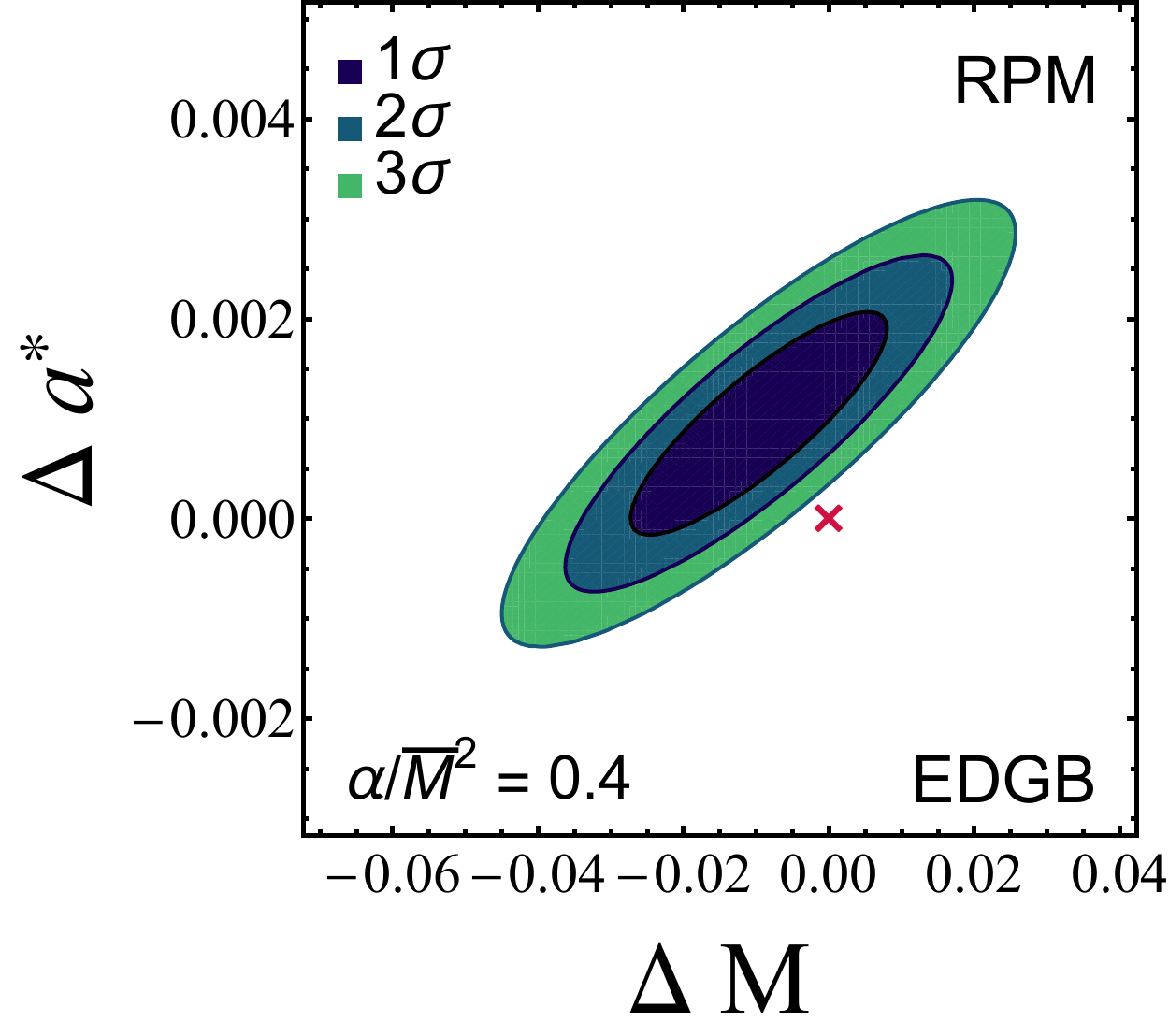}
\includegraphics[width=5.3cm]{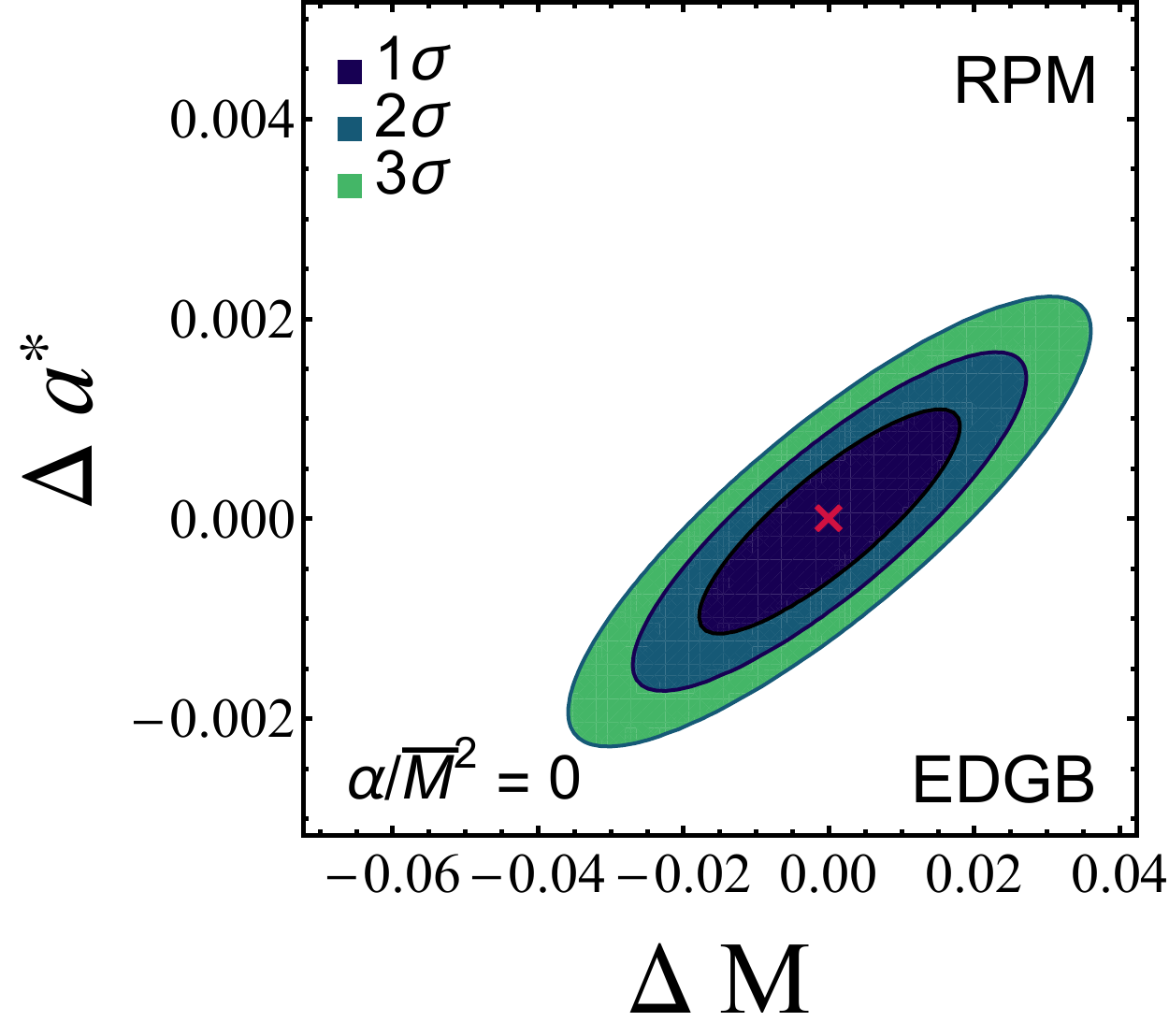}\\
\includegraphics[width=5.3cm]{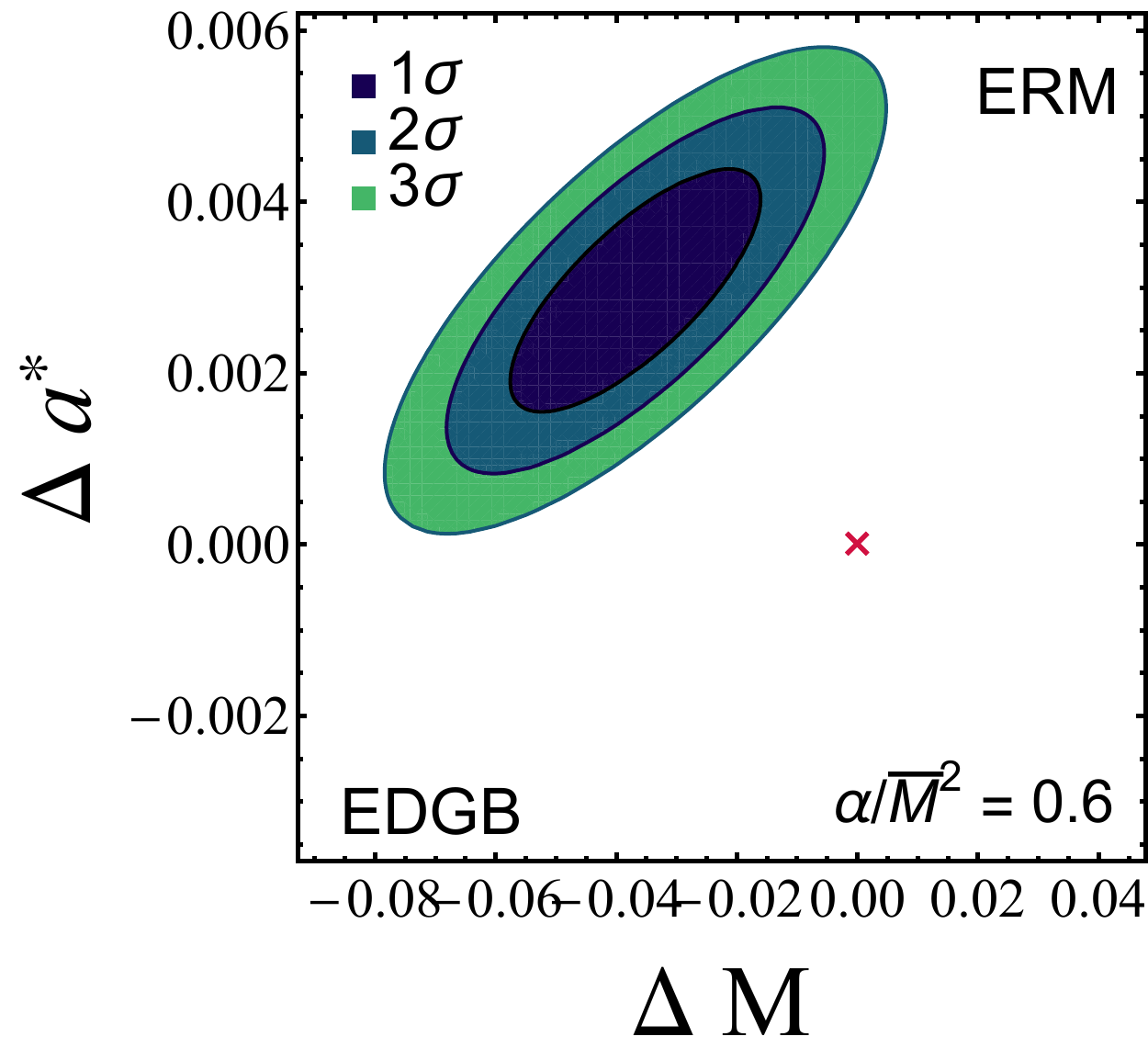}
\includegraphics[width=5.3cm]{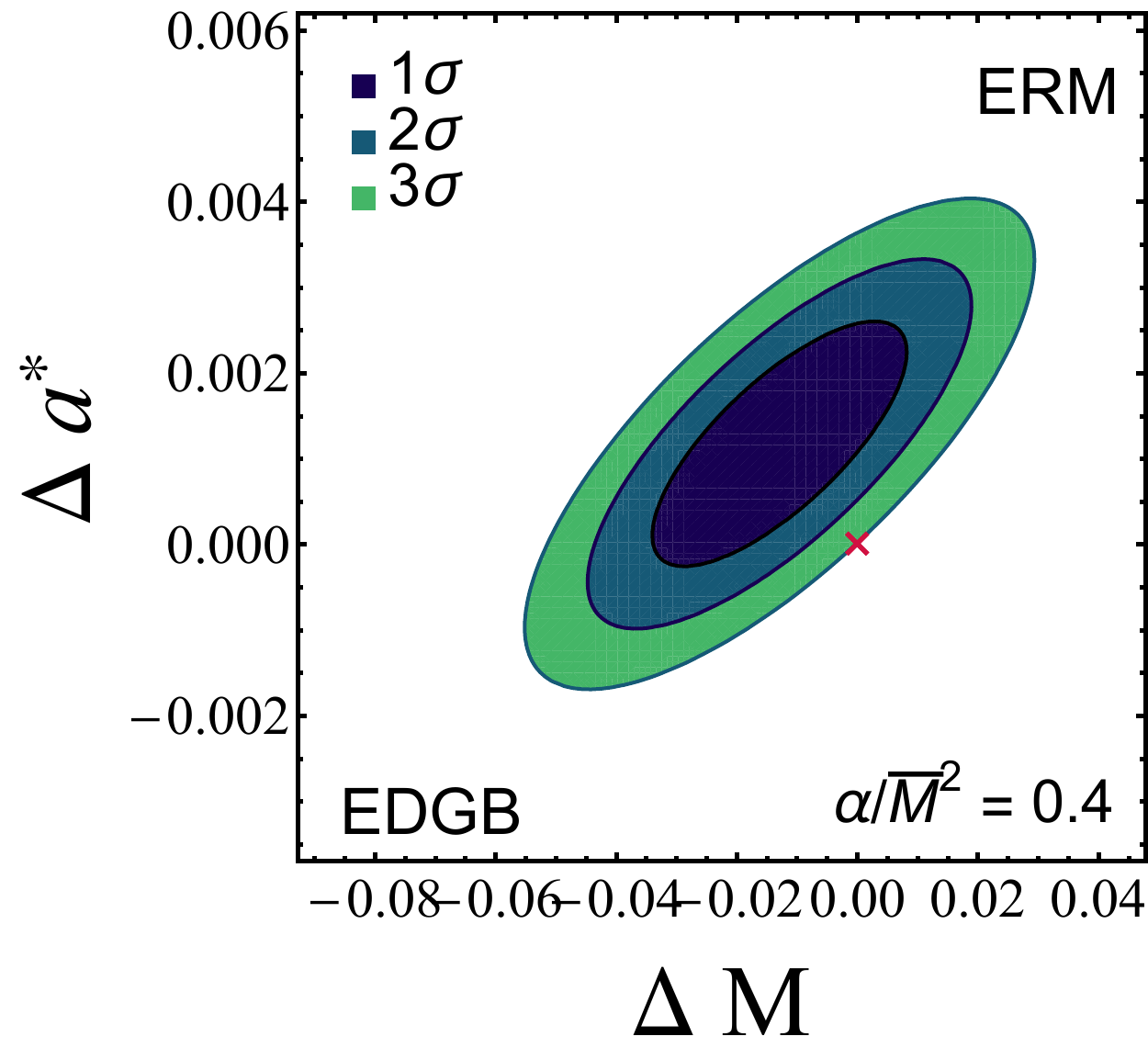}
\includegraphics[width=5.3cm]{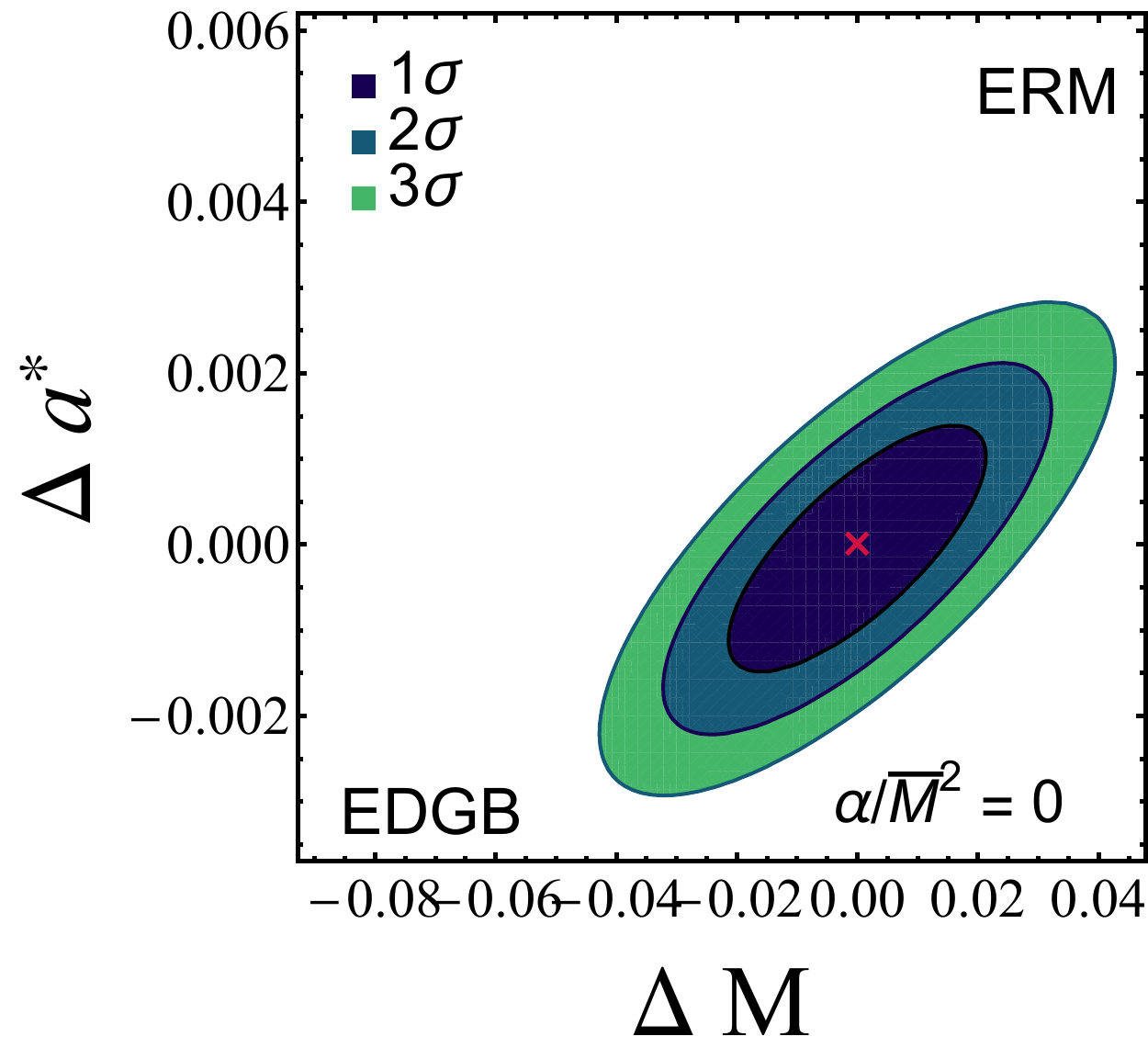}
\caption{Confidence levels in the $(\Delta M,\Delta a^\star)$ plane, obtained
for the RPM and 
the ERM model, by using geodesic frequencies from EDGB gravity.
The red crosses identify the origin of the plane. All data refer to a BH with
mass
$\bar{M}=5.3M_{\odot}$, spin $\bar{a}^\star=0.5$, and different coupling
parameters
$\alpha/\bar{M}^2$. (Top row) Center panel refers to the threshold value of
$\alpha/\bar{M}^2$ for
which we can exclude, at $3$-$\sigma$ confidence level, that the two observed
triplets are generated by a Kerr BH.
The right plot corresponds to the GR case for which $\alpha/\bar{M}^2=0$.
(Bottom row) Same as on the top line, but for the ERM.}
\label{ellipses}
\end{figure*}
%%%%%%%%%%%%%%%%%%%%%%%%%%%%%%%%%%%%%%%%%%%%

In order to assess whether these distributions are consistent with 
$M_1=M_2$ and $a^\star_1= a^\star_2$, we define the three variables 
\begin{equation}
\Delta M=M_1-M_2\ ,\ \Delta a ^\star= a^\star_1-a^\star_2\ ,\ \Delta r=r_1-r_2
\end{equation}
checking that the normal distribution ${\cal N}(\vec{\mu},\Sigma)$, with 
$\vec{\mu}=(\Delta M,\Delta a^\star, \Delta r)$ and $\Sigma=\Sigma_1+\Sigma_2$, 
is consistent with a Gaussian of zero mean. Given the $\chi^2$ 
variable
\begin{equation}
\chi^2=(\vec{x}-\vec{\mu})^\tn{T}\Sigma^{-1}(\vec{x}-\vec{\mu})\ ,
\end{equation}
the values of $\chi^2=\delta$, define the confidence levels (CL) of 
$(\Delta M, \Delta a^\star, \Delta r)$ for a specific choice of $\delta$.
We repeat this analysis assuming both the RPM and the ERM to 
interpret the QPOs, assuming the observed triplet as given by 
$(\nu_{\phi},\nu_\tn{nod},\nu_\tn{per})_{i=1,2}$ and $(\nu_{\theta},\nu_-,\nu_\tn{nod})_{i=1,2}$, 
respectively. In the RPM the location of the emission 
radii can be, in principle, chosen freely; following \citep{Maselli:2014fca} we adopt $r_1=1.1r_\tn{Isco}$ and $r_2=1.4r_\tn{Isco}$, 
which give rise to QPO frequencies comparable to those observed in GRO J1655-40.
(Somewhat different choices of the radial coordinates, as long as they are $\leq 2 r_\tn{Isco}$, do not alter significantly our results).
In the ERM approach, instead, for each resonance there is only one radius that satisfies 
condition (\ref{resonancecon}). Specifically, we choose $r_1$ and $r_2$ in order 
to obtain $\nu_{\theta}/\nu_-=3/2$ and $\nu_{\theta}/\nu_-=5/2$.

\begin{figure}[ht]
\centering
\includegraphics[width=4.2cm]{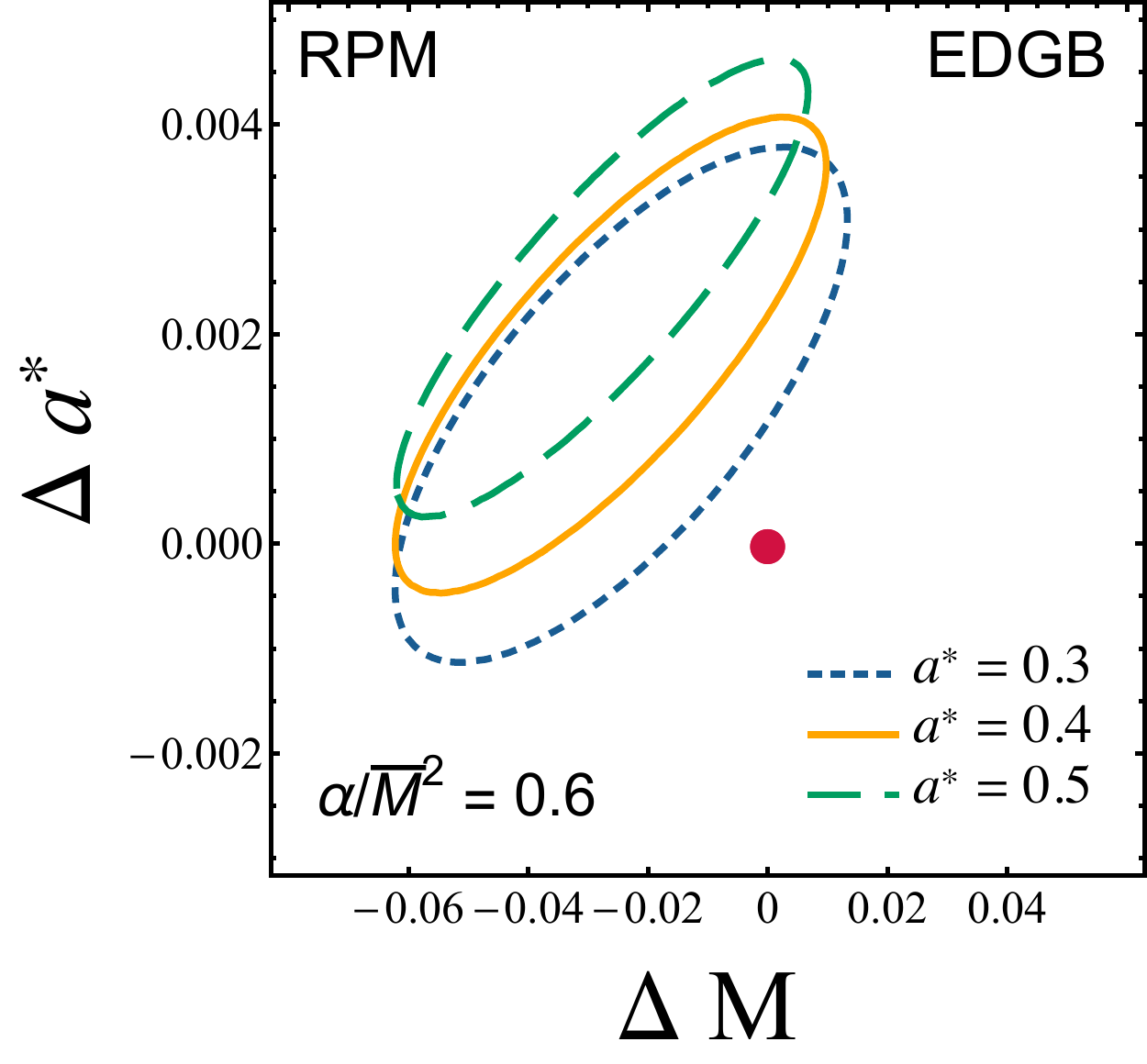}
\includegraphics[width=4.2cm]{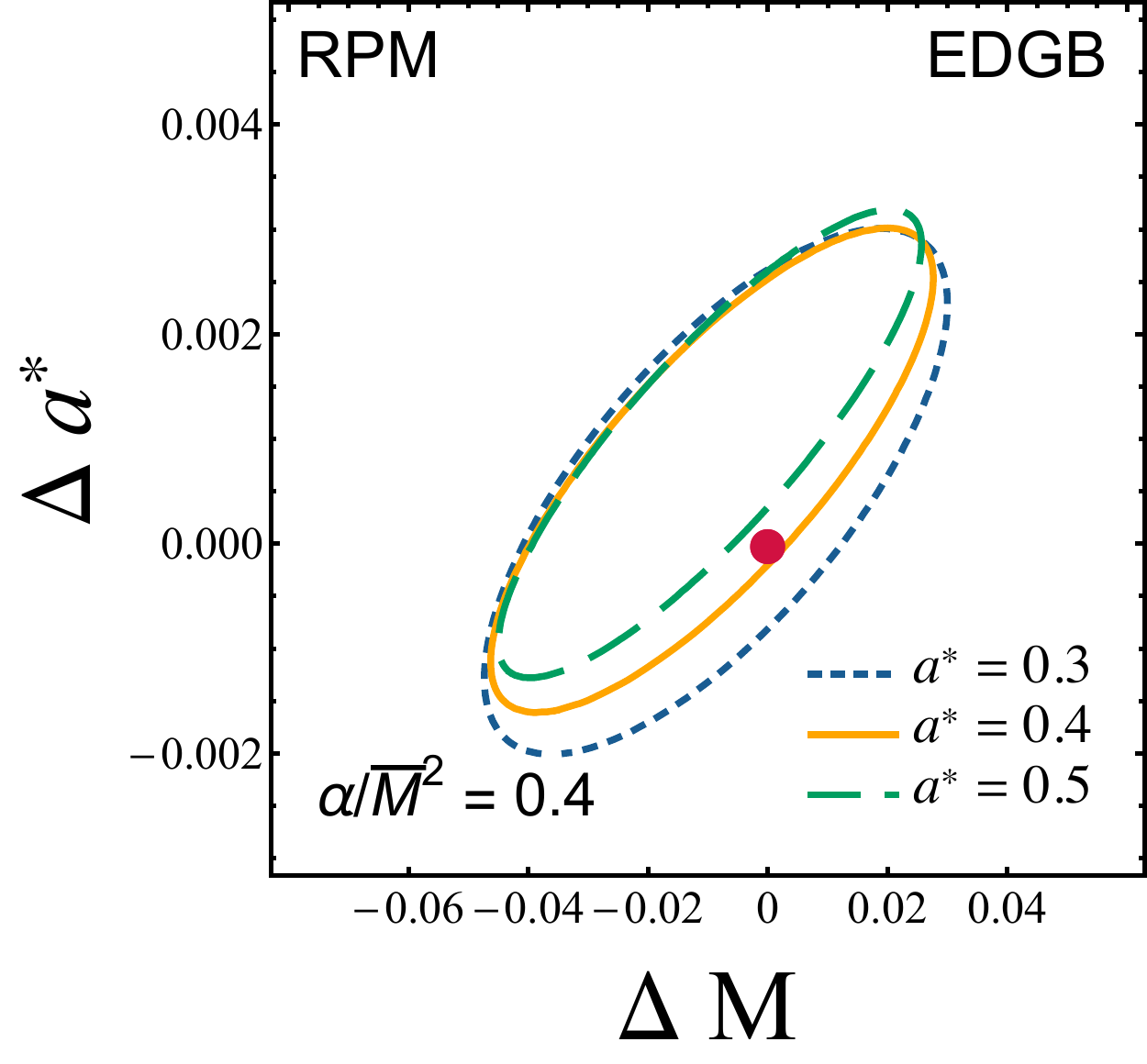}
\caption{$3\sigma$ confidence levels for BHs with spin parameters $\bar{a}^\star=(0.3,0.4,0.5)$ 
and coupling $\alpha/\bar{M}^2=(0.4,0.6)$, using the RPM.}
\label{RPMellipses2}
\end{figure}

The top panels of Fig.~\ref{ellipses} show the CL, at 
$1\sigma$, $2\sigma$ and $3\sigma$ 
in the  $(\Delta M,\Delta a^\star)$ parameter space, 
for BHs with $\bar{a}^\star=0.5$ and $\alpha/\bar{M}^2=(0.6,0.4,0)$, in the 
case in which the QPO frequencies are described by the RPM. 
The red cross identifies the origin of 
the plane, and corresponds to the null hypothesis, for which the two samples 
of QPO triplets derive from the same distribution, i.e. also the input data are 
generated from the Kerr metric. The panel on the right shows the case
$\alpha/\bar{M}^2=0$ as consistency check of our method; it is apparent that 
the three CLs are  
centered on $(\Delta M,\Delta a^\star)=(0,0)$, and thus that the data are consistent with 
$M_1=M_2$ and $a^\star_1=a^\star_2$. The left and central plots refer to BH configurations with 
$\alpha/\bar{M}^2 > 0$: for these models $\Delta M$ and $\Delta a^\star$ are 
both incompatible with 0. In particular, the central upper panel of Fig.~\ref{ellipses} shows that  
if the coupling parameter is $\alpha/\bar{M}^2=0.4$, the data are already incompatible with 
the Kerr metric at more than the $3\sigma$ level. 
Higher BH spins ($a^\star > 0.5$)  would exclude even smaller values of the coupling parameter.
To clarify this point, we carry out the same analysis by varying the BH spin. 
Figure~\ref{RPMellipses2} shows the $3\sigma$ CL for $\alpha/\bar{M}^2=(0.6,0.4)$ 
and $a^\star=(0.3,0.4,0.5)$. For a fixed value of the EDGB coupling 
parameter, increasing BH spins make simulated QPO frequencies
depart more and more from GR predictions.

%%%%%%%%%%%%%%%%%%%%%%%%%%%%%%%%%%%%%%%%%%%%
%%%%%%%%%%%%%%%%%%%%%%%%%%%%%%%%%%%%%%%%%%%%

The bottom panels of Fig.~\ref{ellipses} show the CLs for the same 
BH configurations described above and  QPO frequencies 
calculated in accordance with the ERM.
%%%%%%%%%%%%%%%%%%%%%%%%%%%%%%%%%%%%%%%%%%%%
\begin{figure}[ht]
\centering
\includegraphics[width=4.2cm]{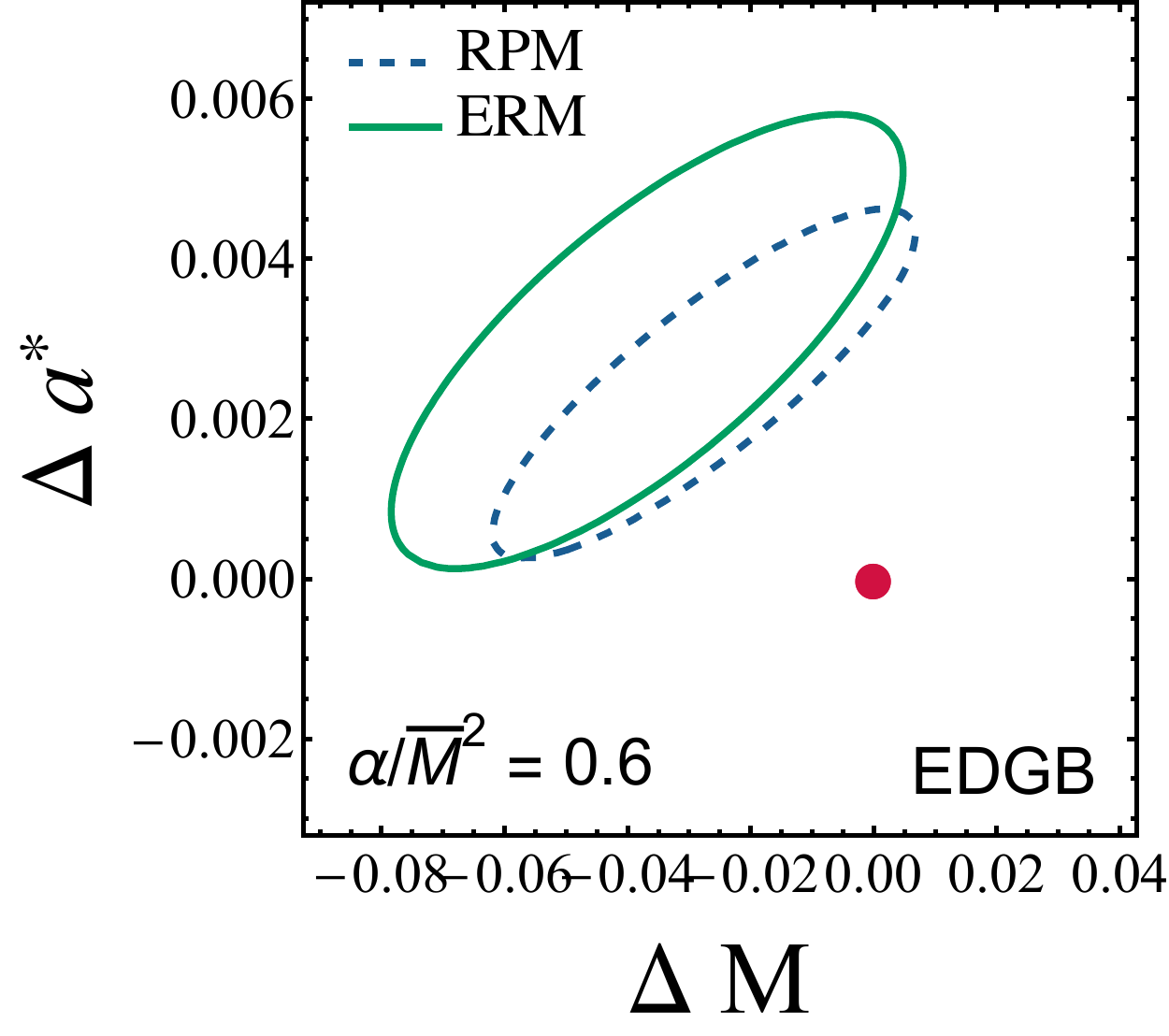}
\includegraphics[width=4.2cm]{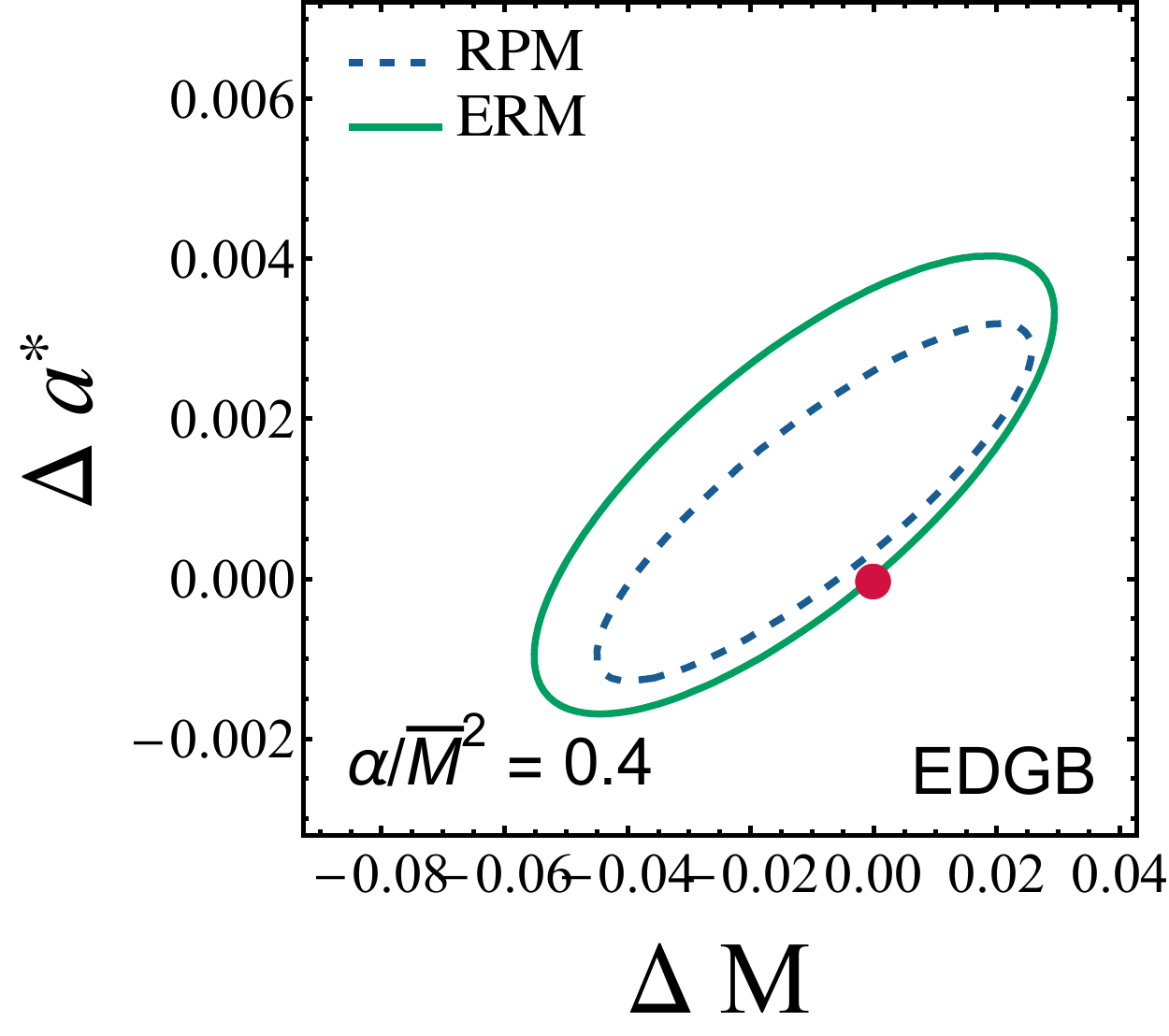}
\caption{$3\sigma$ confidence levels for BHs with spin parameter
$\bar{a}^\star=0.5$
and coupling $\alpha/\bar{M}^2=(0.4,0.6)$, computed for the RPM (dashed
ellipses) and
the resonance model (solid ellipses).}
\label{RPMvsRES}
\end{figure}
%%%%%%%%%%%%%%%%%%%%%%%%%%%%%%%%%%%%%%%%%%%%
The central panel is for $\alpha/\bar{M}^2\sim0.4$, {\it i.e.} the  value 
excluded at slightly more than $3\sigma$ CL also in this case.  In general, the
two frameworks provide similar constraints, as shown in Fig.~\ref{RPMvsRES}, in
which we make a direct comparison between the two QPO models. We note that the
ellipses obtained for the ERM are slightly larger than those computed for the
RPM, irregardless of the EDGB coupling parameter.  
We have also computed the ellipses for RPM and ERM choosing the same $r_2$, 
but the difference between the two approaches still persists.

%%%%%%%%%%%%%%%%%%%%%%%%%%%%%%%%%%%%%%%%%%%%
\begin{figure}[ht]
\centering
\includegraphics[width=4.3cm]{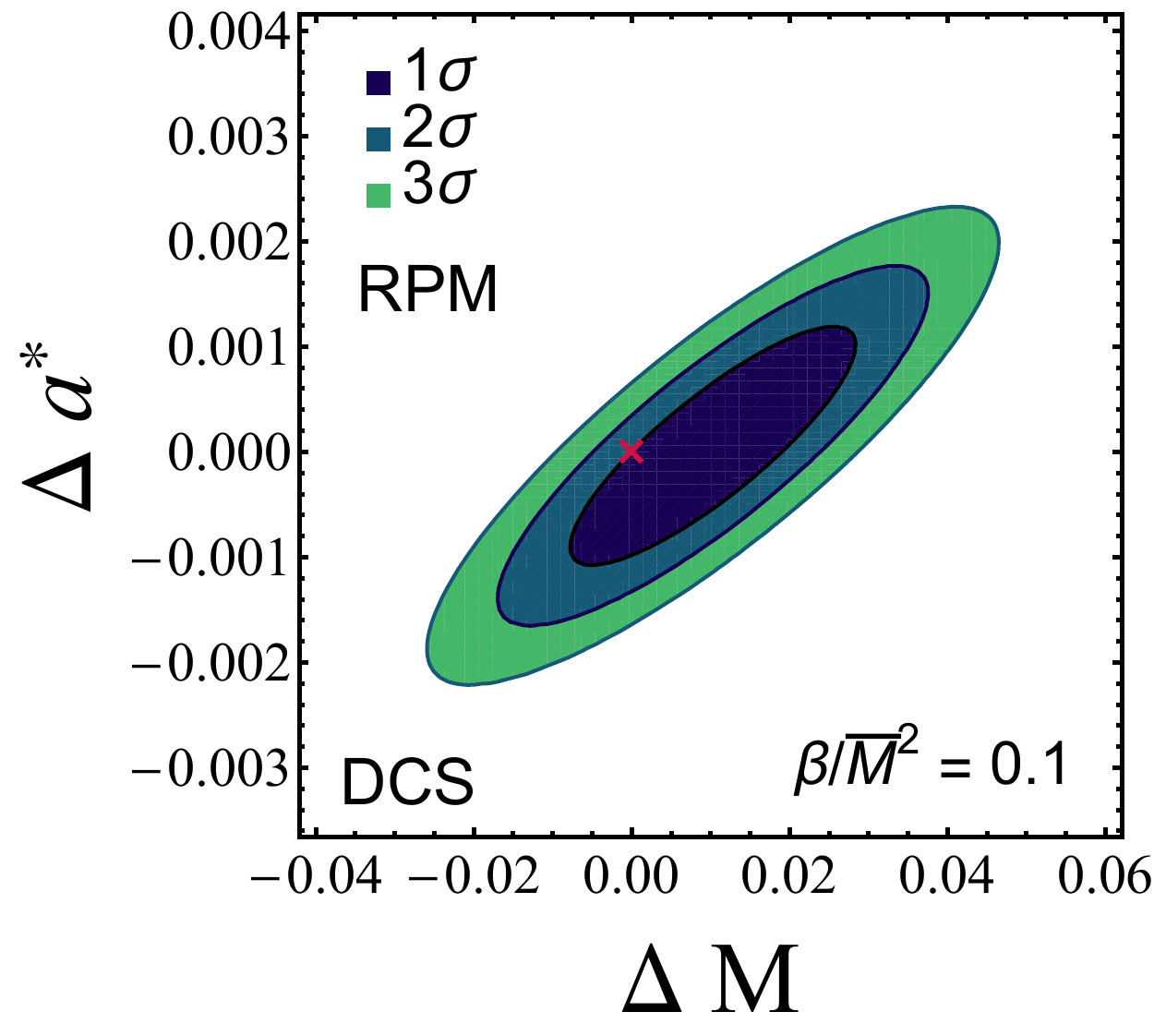}
\caption{Confidence levels computed for the RPM, for a $5.3 M_\odot$ BH with
spin
$\bar{a}^\star=0.5$ and DCS coupling parameter $\beta/\bar{M}^2=0.1$.}
\label{DCSellipse}
\end{figure}
%%%%%%%%%%%%%%%%%%%%%%%%%%%%%%%%%%%%%%%%%%%%

%%%%%%%%%%%%%%%%%%%%%%%%%%%%%%%%%%%%%%%%%%%%
\begin{figure}[ht]
\centering
\includegraphics[width=4cm]{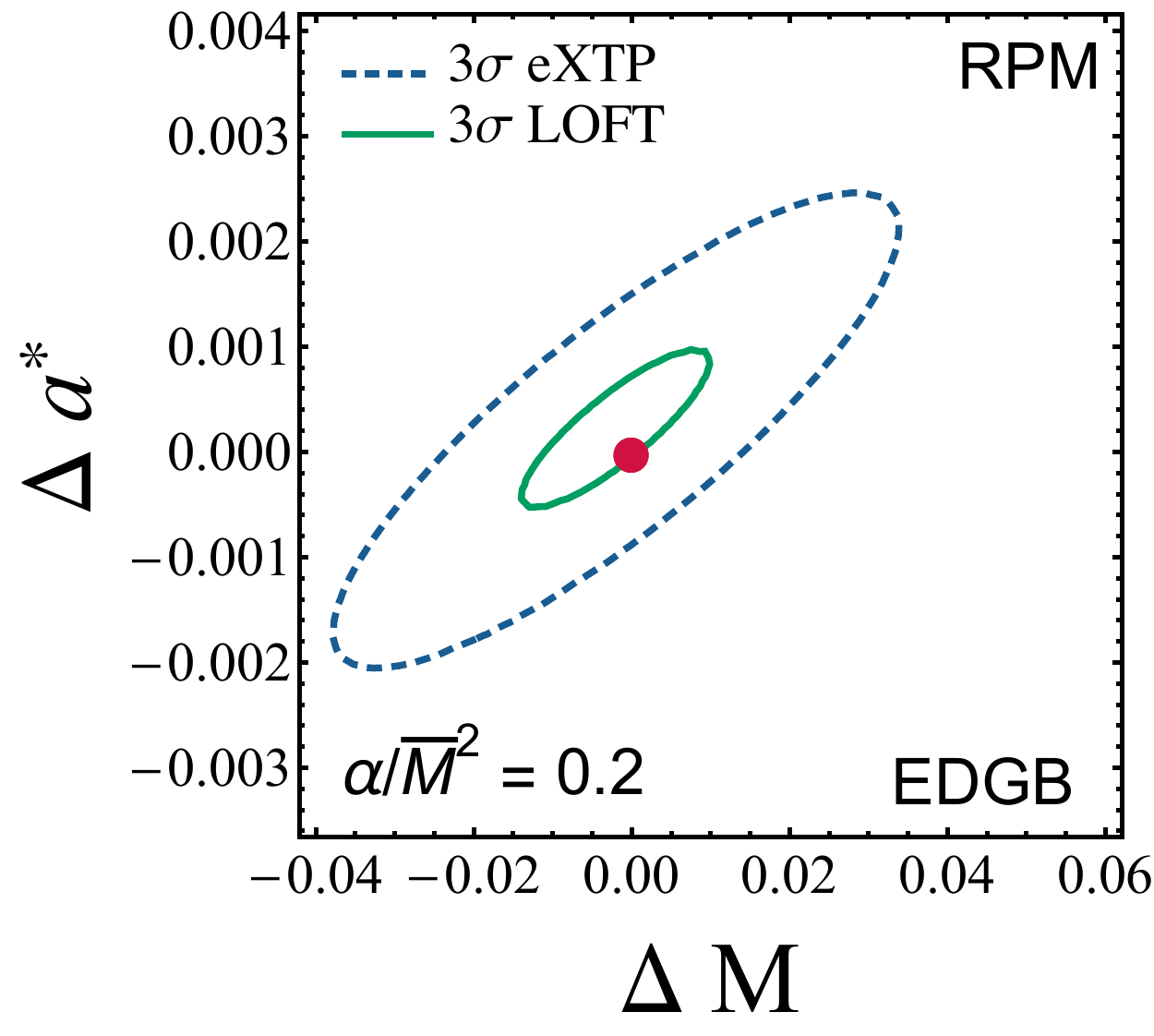}
\includegraphics[width=4cm]{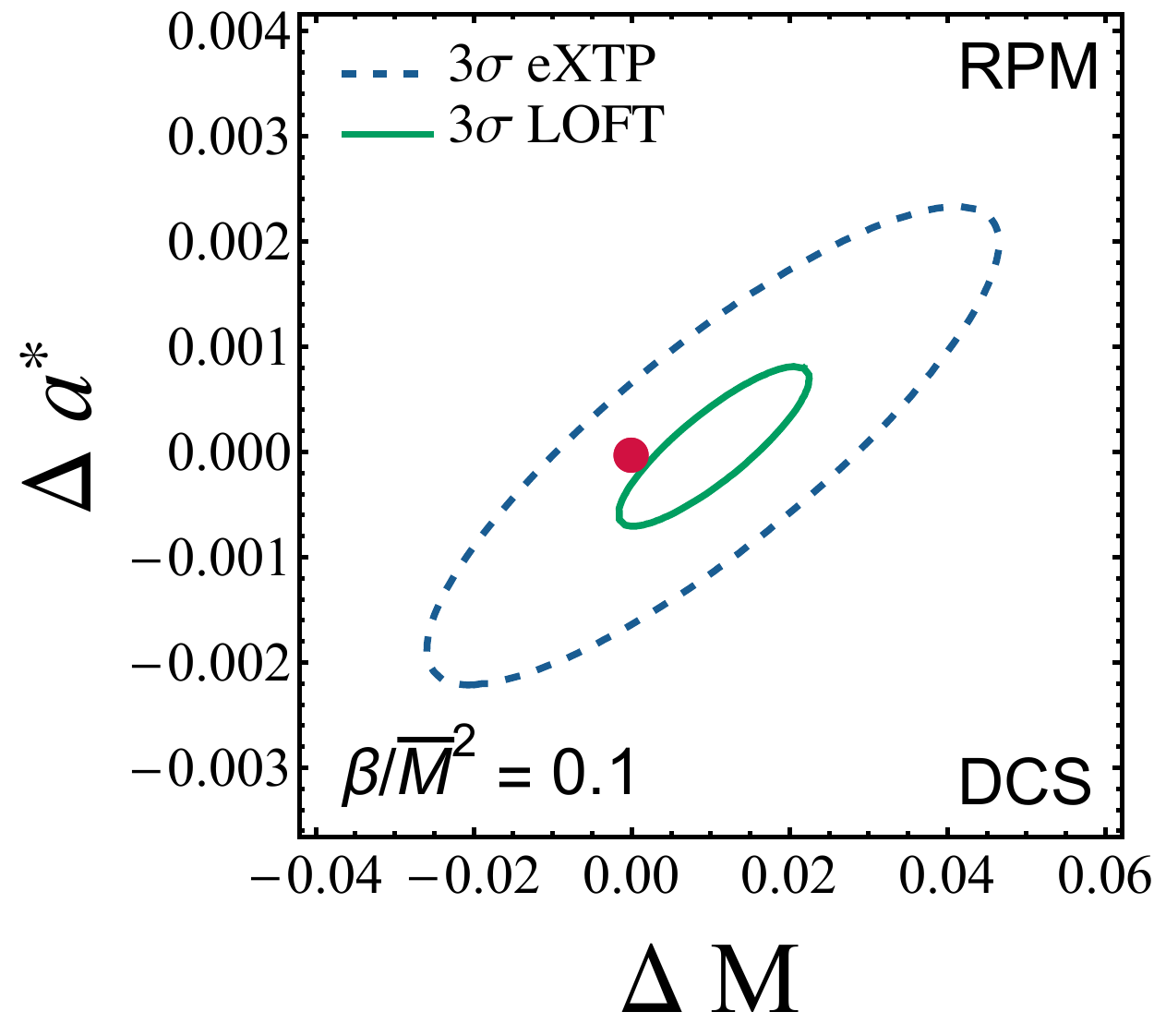}
\caption{$3-\sigma$ confidence levels computed assuming the RPM, for a BH with
spin
$\bar{a}^\star=0.5$, for EDGB (left), and DCS (right). The dashed and solid
contours
refer to a eXTP and LOFT-like experiment, respectively.}
\label{LOFTellipse}
\end{figure}
%%%%%%%%%%%%%%%%%%%%%%%%%%%%%%%%%%%%%%%%%%%%

Much different results are obtained when azimuthal and the epicyclic frequencies
are generated by using DCS theory. The results of our analysis are shown in
Fig.~\ref{DCSellipse}, for a BH with the same mass and spin considered before
(i.e. $\bar{M}=5.3 M_\odot$, $a^\star=0.5$), and DCS coupling parameter
$\beta/\bar{M}^2=0.1$. We emphasize that the latter value is the highest compatible with
a small-coupling approximation of an effective theory of gravity, free of
pathologies.  It is apparent that the cross corresponding to the null
hypothesis (i.e., GR-Kerr frequencies apply) lies between  $2-$ and $3-\sigma$
CL. Therefore, in this case, we would not be able to set statistically
significant constraints on the theory, even for the maximum value
$\beta/\bar{M}^2=0.1$.

Finally, we consider the outcome of our study, when QPO measurements 
with LOFT-type S/N (as in \citep{Maselli:2014fca}) 
are also considered. The two plots of Fig.~\ref{LOFTellipse} show the 
$3-\sigma$ ellipses computed for the RPM in EDGB and DCS and 
for both eXTP-type and LOFT-type QPO data (for the effective area of
LOFT/LAD we use $\times 15$ the effective area of RXTE/PCA, as in \citep{Maselli:2014fca}).
In both gravity theories, a higher S/N 
leads to a significant improvement of the excluded parameter space. This is especially 
evident for DCS, which can now be constrained even for  
$\beta/\bar{M}^2=0.1$. 

We note that solar system experiments based on measurements of the geodesic precession 
and tests of the Newton's law, have already constrained the DCS coupling to 
$\sqrt{\beta}<\mathcal{O}(10^8\tn{km})$ \citep{AliHaimoud:2011fw}, which is much 
weaker than the requirement $\beta/M^2\ll 1$ for $M\ll10^8M_\odot$, i.e. for the 
stellar mass sources considered in this paper. For EDGB gravity, the best bound on the coupling 
parameter comes from observations of low mass X-ray binaries, leading to $\alpha\lesssim 47M_\odot^2$ 
\citep{Yagi:2012gp}, which is looser than the theoretical bound~(\ref{EDGBconst}) for $M\lesssim 8.2M_\odot$.

%%%%%%%%%%%%%%%%%%%%%%%%%%%%%%%%%%%%%%%%%%%%
\section{Conclusion}\label{sec:conclusion}
%%%%%%%%%%%%%%%%%%%%%%%%%%%%%%%%%%%%%%%%%%%%

X-ray QPOs emitted by accreting BHs represents a very promising tool 
to investigate stationary spacetimes in a genuine strong-field and high-curvature regime, 
even though their interpretation is still open to different possibilities.
Future instrumentation such as the eXTP/LAD holds the promise not only to shed light on the origin  
of QPOs, but also to constrain modified gravity theories in extreme astrophysical environments. 
 
In this paper, we have extended the analysis presented in \citep{Maselli:2014fca} in 
several directions: (1) by considering DCS gravity
in addition to EDGB gravity; in an effective-field-theory approach, these are the only quadratic gravity theories admitting BH solutions other than Kerr; (2) by calculating accurate geodesic frequencies for higher 
values of BH spin, up to $a^\star = 0.5$ (as opposed to 0.2); (3) by adopting 
two different geodesic QPO models, the RPM and the ERM.
Most of our simulations here were carried out adopting the S/N expected for 
an X-ray instrument of effective area ($\sim 3.5\ \rm m^2$), comparable to that envisaged for
the eXTP/LAD instrument (which is about a factor of $\sim 6$ larger than that 
of the RXTE/PCA). This is at variance with the simulations in \citep{Maselli:2014fca} 
which used instead a larger effective ($\sim 9\ \rm m^2$ close to that being studied 
for LOFT). 

Our results can be summarized as follows.

\begin{enumerate}
\item Both the RPM and the ERM provide viable frameworks to 
test alternative gravity theories to GR, through high-precision X-ray
measurements of QPOs.
\item Even for moderately fast-rotating BHs, with spin parameter $a^\star\sim0.5$, an eXTP-type mission 
may set stringent constraints on the EDGB coupling parameter.
\item As already noted in \citep{Maselli:2014fca}, our ability to distinguish between GR and EDGB 
increases with the spin of the accreting object. By extrapolating our results here we suggest that 
maximally spinning  BHs, with $a^\star\sim1 $, are the best probes of quadratic gravity.
\item With an eXTP-type mission,  none of the models considered here set useful bounds on DCS theory. 
This is mainly due to the small-coupling parameter ($\leq 0.1$), which is consistent with the requirement 
that DCS be an effective theory of gravity, and to the fact that DCS gravity introduces smaller corrections 
to GR BH geometries.
\item A LOFT-like observatory, with its improved S/N, would constrain more tightly  the 
parameter space of both modified theories we have considered. In particular, this would allow us 
to set new bounds on DCS gravity, which would otherwise be insensitive to the QPOs diagnostic.
\end{enumerate}

Our study shows that the application of geodesic models to future
high S/N QPO measurements holds the potential to test alternative gravity theories in the strong-field 
regime, with the results appearing to be especially promising for EDGB theory.

%%%%%%%%%%%%%%%%%%%%%%%%%%%%%%%%%%%%%%%%%%%%
%%%%%%%%%%%%%%%%%%%%%%%%%%%%%%%%%%%%%%%%%%%%

\section*{Acknowledgments}
\vskip 2cm
P.P. acknowledges support from FCT-Portugal through the project IF/00293/2013.
L.S. Acknowledges funding from the ASI-INAF contract I/004/11/1. This work was supported 
by the H2020-MSCA-RISE-2015 Grant No. StronGrHEP-690904 and by the COST action CA16104 
''GWverse".

%%%%%%%%%%%%%%%%%%%%%%%%%%%%%%%%%%%%%%%%%%%%

\bibliography{bibnote}
\bibliographystyle{apj}

\end{document}